\documentclass[trackchanges,twocolumn]{aastex701}
\usepackage{anyfontsize}
\usepackage{graphicx}	
\usepackage{amsmath}	

\newcommand{\vzfour}{\langle \tilde{v}_{los}^{4} \rangle}
\submitjournal{ApJ}

\begin{document}

\title[Dynamical Mass Profiles of Galaxy Groups]{The Dynamical Mass Profiles of Galaxy Groups - Application to IllustrisTNG-300}

\correspondingauthor{Rita Tojeiro}

\author[orcid=0000-0002-6592-9916,gname=Alfie,sname=Russell]{Alfie Russell}
\affiliation{SUPA, School of Physics and Astronomy, University of Saint Andrews, North Haugh, Saint Andrews, Fife KY16 9SS, UK}
\email[show]{apr8@st-andrews.ac.uk}  

\author[orcid=0000-0001-5191-2286,gname=Rita, sname=Tojeiro]{Rita Tojeiro} 
\affiliation{SUPA, School of Physics and Astronomy, University of Saint Andrews, North Haugh, Saint Andrews, Fife KY16 9SS, UK}
\email[show]{rmftr@st-andrews.ac.uk}

\author[orcid=0000-0002-5908-6852,gname=Anne-Marie, sname=Weijmans]{Anne-Marie Weijmans} 
\affiliation{SUPA, School of Physics and Astronomy, University of Saint Andrews, North Haugh, Saint Andrews, Fife KY16 9SS, UK}
\email{amw23@st-andrews.ac.uk}


\begin{abstract}
We present a new rescaling and stacking method to test cold dark matter predictions through the dynamical modelling of galaxy groups using the Jeans equation. We test the viability of this method on idealised, line-of-sight group data at $z=0$ in the IllustrisTNG-300 simulation, using subhalos and satellite galaxies as tracers. Dynamical mass profiles are measured using the code `\textsc{gravsphere}' for seven distinct quality cuts, three on the dark matter substructure and four on the `visible' satellite galaxies. When using the dark matter substructure, we find that selecting subhalos by their infall time onto the group provides the best recovery of the true mass profile. When using satellite galaxies we instead find that selecting quenched satellites provides the best recovery. In both cases the mass profile is recovered within $<1\sigma$ of the true value up to $\sim 1.5$ times the median host virial radius. However, we also find the recovered velocity anisotropy to be biased towards negative values for all seven of our subhalo samples beyond the median host virial radius. This suggests dynamical systematics may still be present in the data, which will need to be quantified in future work before application to surveys.
\end{abstract}



\section{Introduction}
All massive systems exhibit a ``missing mass" problem, where their visible baryonic content is insufficient to explain the properties we observe. This was first observed in the velocity dispersions of galaxy clusters \citep{Zwicky_1933,Zwicky_1937} and is more famously known through the flat rotation curves of galaxies \citep{Rubin_1970, Sofue_2001}. The standard solution is to introduce a cold dark matter component (CDM) that dominates the dynamics of these systems, as in the current cosmological model $\Lambda$CDM \citep{Efstathiou_1990,Ostriker_1995}. $N$-body cosmological simulations predict this CDM to form halos that conform to the self-similar NFW profile, \citep{NFW1996, NFW_1997} and so measuring the mass profiles of bound systems can be used to test the nature of the ``missing mass".

The NFW profile is characterised by a `cuspy' inner density of $\rho(r) \propto r^{-1}$, but this is contradicted by observations of dwarf galaxies \citep{Flores_1994, Moore_1994,Oh_2015}, which prefer a cored dark matter profile with $\rho(r) \propto r^{0}$ near the centre of the halo. Proposed solutions to this `cusp-core problem' include strong baryonic feedback effects \citep{Lazar_2020}, alternative dark matter models such as Warm Dark Matter \citep[WDM; ][]{Bode_2001}, Fuzzy Dark Matter \citep[FDM; ][]{Hu_2000}, and Self-interacting Dark Matter \citep[SIDM; ][]{Spergel_2000}, and even modified gravity theories such as Modified Newtonian Dynamics \citep[MOND; ][]{Milgrom_1983}, Scalar-Tensor-Vector Gravity \citep[MOG; ][]{Moffat_2006} or Emergent Gravity \citep[EG; ][]{Verlinde_2017}. These different solutions are difficult to differentiate between using only galaxies precisely because they are designed to describe these systems. Baryonic feedback effects on these scales can further confuse what they even \textit{should} predict. This motivates investigating the outer regions of halos or larger mass halos, where baryonic effects are negligible and the NFW profile predicts an outer density slope of $\rho(r) \propto r^{-3}$.

Measuring these outer profiles in galaxy halos is currently only possible using galaxy-galaxy weak lensing. Recent studies \citep{Brouwer_2021, Mistele_2024_RAR, Mistele_2024} seem to dispute NFW expectations and find a shallower decline of $\rho(r) \propto r^{-2}$. This is more in line with modified gravity theories such as MOND, MOG and EG, which are designed to produce the `infinitely flat rotation curves' that this density profile corresponds to.

The exact opposite scenario is observed in larger mass cluster halos, where many more methods are available. Dynamical studies using the Jeans equation \citep[e.g.][]{Lokas_2003_4thlos,Katgert_2004,Li_2023} or the caustic method \citep[e.g.][]{Biviano_2003,Rines_2003,Rines_2013} find consistency with NFW in the outer regions and similar results are seen with gravitational weak lensing \citep[e.g.][]{Umetsu_2011,Sereno_2017,Singh_2025}. The inner regions of cluster halos, where baryonic effects are much less impactful when compared to galaxy halos, also show consistency with NFW. This is observed when using member galaxy dynamics \citep{Carlberg_1997,Mamon_2019}, gravitational lensing \citep{Merten_2015,Umetsu_2016} and when measuring the pressure of hot gas \citep{Arabadjis_2002,Lewis_2003,Eckert_2022} at the centre of clusters \citep[see][for more complete reviews of cluster mass profile determinations]{Pratt_2019,Umetsu_2020}. In fact, MOND-like theories are also known to generically underpredict the total masses of galaxy clusters \citep{Sanders_1999, Pointecouteau_2005,Kelleher_2024}.

This discrepancy between galactic halos and cluster halos raises the question of whether these systems are genuinely physically distinct, or whether there is some methodological issue driving this difference. It is therefore important to extend methods used in clusters to lower mass systems and begin to check for consistency with galaxy-galaxy weak lensing. Galaxy groups represent an intermediate mass between galaxies and clusters, however, they are not massive enough for strong lensing in their outer regions, they do not contain the hot X-ray gas necessary for pressure studies and they typically contain too few member galaxies to apply standard dynamical modelling techniques. 

We aim to overcome this last point and measure dynamical mass profiles for galaxy groups by rescaling and stacking groups. The goals for this paper are to test such a process on idealised data from hydrodynamical simulations, where the true mass profiles are known. This will highlight in what regimes accurate mass profiles may be recovered, what types of quality cuts may be required and whether there are significant systematic effects to expect. If such a methodology is feasible, it could be applied to modern spectroscopic surveys to bridge the mass profile gap between galaxy halos and cluster halos. We will start by detailing the theory of dynamical modelling using the Jeans equation in \autoref{sec:theory}. We will then discuss the preparation of idealised data from the TNG-300 simulations in \autoref{sec:data}, along with the stacking methodology we aim to employ. In \autoref{sec:Methods} we will describe the numerical methods we used to model dynamical mass profiles and our findings will be presented in \autoref{sec:results}. We discuss these results and other considerations needed before application to observations in \autoref{sec:discussion}. We then provide our conclusions in \autoref{sec:conclusion}.

\section{Theory}\label{sec:theory}

We will use the spherical Jeans equation \citep{Binney_1980,Binney_2008} to perform our dynamical modelling analysis. At radius $r$ of a gravitationally bound system, the Jeans equation relates the enclosed mass, $M(<r)$, to the orbits of tracers by
\begin{equation}\label{eq:Jeans 3D}
    \frac{1}{\nu(r)} \frac{\partial}{\partial r}\left[\nu(r) \sigma^{2}_{r}(r) \right] + \frac{2 \beta(r) \sigma^{2}_{r}(r)}{r} = -\frac{G M(<r)}{r^{2}}.
\end{equation}
$\sigma_{r}^{2}$ is the radial velocity dispersion of the tracers, $\nu$ is the number density of the tracers and $\beta$ is the velocity anisotropy, given by
\begin{equation}\label{eq:Beta definition}
    \beta(r) = 1 - \frac{\sigma^{2}_{t}}{2\sigma^{2}_{r}} = \frac{3}{2} - \frac{\sigma^{2}_{tot}}{2\sigma^{2}_{r}}.
\end{equation}
We define the tangential velocity dispersion as $\sigma^{2}_{t} \equiv \sigma^{2}_{\theta} + \sigma^{2}_{\phi}$, where $\theta$ and $\phi$ denote the polar and azimuthal directions respectively, and the total velocity dispersion as $\sigma_{tot}^{2} = \sigma_{r}^{2} + \sigma_{\theta}^{2} + \sigma_{\phi}^{2}$. \autoref{eq:Jeans 3D} is satisfied for steady-state, spherically symmetric systems in which the tracers are in dynamical equilibrium with no systematic rotation.

It is often more useful to solve for the radial velocity dispersion, which is found by integrating \autoref{eq:Jeans 3D} \citep{van_der_Marel_1994} such that
\begin{equation}\label{eq:Jeans 3D integrated}
    \sigma_{r}^{2}(r) = \frac{1}{\nu(r)g(r)}\int_{r}^{\infty} \frac{GM(<r^{\prime})\nu(r^{\prime})}{\left(r^{\prime}\right)^{2}}g(r^{\prime})dr^{\prime},
\end{equation}
where $g(r)=\exp\left[2\int \frac{\beta(r)}{r}dr\right]$. 

However, \autoref{eq:Jeans 3D integrated} requires 3-D knowledge of the system, whereas observations are limited to line-of-sight and projected measurements. \citet{Binney_1982} show that the corresponding line-of-sight velocity dispersion is
\begin{equation}\label{eq:Jeans 1-D}
    \sigma_{los}^{2}(R)=\frac{2}{\Sigma(R)} \int_{R}^{\infty} \left(1 - \beta(r) \frac{R^{2}}{r^{2}} \right) \frac{\nu(r)\sigma_{r}^{2}(r)r}{\sqrt{r^{2} - R^{2}}} \mathrm{d}r,
\end{equation}
with $\Sigma(R)$ representing the surface density of tracers at projected separation $R$. $\sigma_{los}^{2}$, $\Sigma$ and $R$ are all directly observable, while $\nu$ can be found using the inverse Abel transform as
\begin{equation}\label{eq:nu integral}
    \nu(r) = -\frac{1}{\pi}\int_{r}^{\infty} \frac{d\Sigma}{dR} \frac{dR}{\sqrt{R^{2} - r^{2}}}.
\end{equation}
This leaves only $\beta$ and $M(<r)$ as free parameters in \autoref{eq:Jeans 1-D} (through their effect on $\sigma_{r}^{2}$ from \autoref{eq:Jeans 3D integrated}). However, they are entirely degenerate meaning the mass profile cannot be constrained with \autoref{eq:Jeans 1-D} alone.

\citet{Merrifield_1990} showed this degeneracy can be broken by including fourth order velocity moments, $\langle v_{los}^{4} \rangle$. They derived two `virial shape parameters',
\begin{equation}\begin{aligned}\label{eq:4th order beta 1}
    v_{s1} &= \frac{2}{5}\int_{0}^{\infty}GM\nu\left(5 - 2\beta \right) \sigma_{r}^{2}r\mathrm{d}r \\
    & = \int_{0}^{\infty} \Sigma \langle v_{los}^{4} \rangle R \mathrm{d}R,
\end{aligned}\end{equation}
\begin{equation}\begin{aligned}\label{eq:4th order beta 2}
    v_{s2} &= \frac{4}{35}\int_{0}^{\infty}GM\nu\left(7 - 6\beta \right) \sigma_{r}^{2}r^{3}\mathrm{d}r \\
    & = \int_{0}^{\infty} \Sigma \langle v_{los}^{4} \rangle R^{3} \mathrm{d}R,
\end{aligned}\end{equation}
which independently relate the line-of-sight and radial information of the system. Applying these extra constraints to data allow both $\beta$ and $M(<r)$ to be determined from line-of-sight data only.

We employed the codes \textsc{binulator} and \textsc{gravsphere} \citep{Gravsphere_1, Gravpshere_2, Gravsphere_3, Gravsphere_binulator} to apply the Jeans equations to simulated galaxy groups from the IllustrisTNG simulations, where the complete information of the system is known. We will describe \textsc{binulator} and \textsc{gravsphere} later in \autoref{sec:Methods}.

\section{Data}\label{sec:data}

\subsection{TNG300 groups}\label{subsec:TNG groups}

The IllustrisTNG simulations are a suite of cosmological hydrodynamical simulations ran on the moving-mesh code \textsc{arepo} \citep{Arepo}. A PLANCK 2015 cosmology was assumed for the TNG suite, with $\Omega_{\Lambda,0}=0.6911$, $\Omega_{m,0}=0.3089$, $\Omega_{b,0} = 0.0486$, $\sigma_{8}=0.8159$, $n_{s}=0.9667$ and $h=0.6774$ \citep{PLANCK_2015}.

The TNG simulations cover a range of box sizes and mass resolutions, of which we choose the largest: TNG300-1 (TNG300 moving forwards). TNG300 simulates $2500^{3}$ dark matter particles and $\sim2500^{3}$ gas cells on a periodic box with sides of 205 cMpc$/h$. The mass resolution is $4.0\times10^{7}M_{\odot}/h$ for dark matter particles and $7.6\times10^{6}M_{\odot}/h$ for the gas cells. Further information about TNG300 and the other TNG simulations can be found on the TNG website\footnote{\url{https://www.tng-project.org/data/}} and in \citet{TNG_2018_1, TNG_2018_2, TNG_2018_3, TNG_2018_4, TNG_2018_5, TNG_2019}.

Group catalogs are publicly available for these simulations and were produced using a Friends-of-Friends (FoF) algorithm \citep{FoF_1985} to identify dark matter halos. The \textsc{subfind} algorithm \citep{Subfind_2001} was also applied to each group to identify gravitationally bound substructure (subhalos). We will refer to the most massive subhalo in each group as the `central'. For each group and subhalo these catalogs provide a virial mass ($M_{200}$), a virial radius ($r_{200}$), a position (defined as the coordinate of the particle with the minimum gravitational potential) and a peculiar velocity ($v_{pec}$).

For our sample, we first find the groups at $z=0$ with a virial mass between $10^{11.3}M_{\odot}/h$ and $10^{14}M_{\odot}/h$ which possess at least five subhalos (64888 in total). From these we select 25,000 to analyse, to save on computational time and cost. Groups are sorted by mass in the TNG catalogs, so we select using an even spacing in the group catalog's IDs, which should ensure the analysed subsample is representative of the full sample. The 25,000 selected groups contain 674341 subhalos in total (discounting the centrals).

We limit ourselves to $z=0$ for this proof of concept. The lower mass limit of $10^{11.3}M_{\odot}/h$ ensures the smallest groups contain at least 5000 particles within the virial radius, which is necessary to ensure stable density profiles \citep{Schaller_2015,Sorini_2025}. The upper mass limit of $10^{14}M_{\odot}/h$ roughly corresponds to galaxy clusters, whose mass profiles are already well understood.

\subsection{Rescaling group data}\label{subsec:Rescaling}
Due to the low number of satellite galaxies in individual groups, it is difficult to accurately measure velocity dispersions and fourth-order moments. To be reliable, a single velocity dispersion measurement requires $\gtrsim 10$ satellites, whereas the fourth-order velocity moment requires $\gtrsim 10^{2}$ \citep{Merrifield_1990}. For comparison, a single group of mass $10^{13}M_{\odot}/h$ may contain a few to a few tens of satellite galaxies at most.

The galaxy-galaxy weak lensing studies \citep{Brouwer_2021,Mistele_2024,Mistele_2025} face a similar issue; the weak lensing signal-to-noise ratio (SNR) around individual galaxies is too weak to accurately measure image distortions. They overcome this limitation by stacking data from many galaxies at once. They stack within baryonic mass bins, so that the signal from like galaxies (i.e. similar mass) is preserved.

We aim to improve upon this method by rescaling the data on a group-by-group basis. If done correctly, this should maximise the signal whilst also minimising the difference between the individual rescaled mass profiles. Such a rescaling is theoretically possible as collisionless $N$-body simulations predict dark matter halos to have self-similar density profiles \citep{NFW1996, NFW_1997}, described by the NFW profile
\begin{equation}\label{eq:NFW rho}
    \rho_{NFW}(r)=\frac{\rho_{0}}{\frac{r}{r_{s}} \left(1 + \frac{r}{r_{s}} \right)^{2}}.
\end{equation}
Here, $\rho_{0}$ is a normalisation factor and $r_{s}$ is the scale radius. We can therefore define a rescaled density profile of the form
\begin{equation}\label{eq:rho rescaled NFW}
    \tilde{\rho}_{NFW}(\tilde{r}) \propto \frac{1}{\tilde{r}\left(1+\tilde{r}\right)^{2}},
\end{equation}
where $\tilde{r}=\frac{r}{r_{s}}$. Such a rescaled profile is identical for any NFW halo, so would allow any arbitrary sample of NFW halos to be stacked in this rescaled space.

Modern convention parametrises halos using a radius $r_{\Delta}$, a mass $M_{\Delta}$ and a concentration $c_{\Delta}$ as these are easier to obtain observationally. $r_{\Delta}$ represents the radius within which the average density of the halo is $\Delta$ times the critical density of the universe, $\rho_{c}$. $M_{\Delta}$ is then the mass of this sphere, so is given by
\begin{equation}\label{eq:Virial mass X}
M_{\Delta}=\Delta\rho_{c}\left(\frac{4\pi}{3}r_{\Delta}^{3} \right).
\end{equation}
Finally, the concentration $c_{\Delta}$ relates $r_{s}$ and $r_{\Delta}$ by
\begin{equation}\label{eq:Concentration X}
    r_{s}=\frac{r_{\Delta}}{c_{\Delta}},
\end{equation}
and is interpreted as measuring how centrally concentrated the halo's mass is. Setting $\Delta\simeq200$ for a standard $\Lambda$CDM cosmology means $r_{\Delta}$ roughly corresponds to the radius within which the particles bound to the halo are virialised \citep{Eke_1996}, so $r_{200}$ and $M_{200}$ are often referred to as the virial radius and virial mass respectively. For simplicity, we set $\Delta=200$ so we can use the $M_{200}$ and $r_{200}$ provided by the TNG300 group catalog.

Using \autoref{eq:Concentration X} we see that the change in variable for separations and projected separations now becomes
\begin{equation}\label{eq:r rescale virial}
    \tilde{r} = \frac{r}{r_{s}}, \tilde{R}=\frac{R}{r_{s}}
\end{equation}
when using the virial parameters. For the normalisation, we consider the NFW enclosed mass profile, which has the form
\begin{equation}\label{eq:Mass NFW}
    M_{NFW}(<r) = 4 \pi \rho_{0}r_{s}^{3}\left[\ln\left(\frac{r+r_{s}}{r_{s}}\right) - \frac{r}{r + r_{s}} \right].
\end{equation}
Since $M_{NFW}(<r_{200})=M_{200}$, we can define a scale mass such that 
\begin{equation}
    M_{s} \equiv 4\pi\rho_{0}r_{s}^{3} = \frac{M_{200}}{\ln(1+c_{200}) - \frac{c_{200}}{1+c_{200}}},
\end{equation}
and so the change of variable in mass is
\begin{equation}\label{eq:M rescale virial}
    \tilde{M} = \frac{M}{M_{s}}.
\end{equation}

For the dynamical analysis, we can additionally define a scale time $t_{s}$. Since we generally expect
\begin{equation}
\frac{v^{2}}{r} \propto \frac{GM}{r^{2}},
\end{equation}
we see that an obvious choice is the case where the gravitational constant is rescaled to $\tilde{G}=1$. Dimensional arguments produce
\begin{equation}
    t_{s} \equiv \sqrt{\frac{r_{s}^{3}}{GM_{s}}},
\end{equation}
although this is equivalent to defining a scale velocity of
\begin{equation}
    v_{s} \equiv \sqrt{\frac{GM_{s}}{r_{s}}}.
\end{equation}
The rescaled velocities are then
\begin{equation}\label{eq:v rescale}
    \tilde{v} = v \sqrt{\frac{r_{s}}{GM_{s}}},
\end{equation}
with corresponding uncertainties
\begin{equation}\label{eq:v unc rescale}
    u_{\tilde{v}} = u_{v}\sqrt{\frac{r_{s}}{GM_{s}}}.
\end{equation}

These rescalings yield a predicted rescaled mass profile of
\begin{equation}\label{eq:Mass rescaled}
    \tilde{M}(<\tilde{r}) = \ln(1+\tilde{r}) - \frac{\tilde{r}}{1+\tilde{r}},
\end{equation}
for an NFW system. Notably, this profile is \textit{independent of the mass or concentration} of the original halo, providing a clear prediction for NFW systems. The parameter independence of this profile also allows us to stack \textit{any arbitrary subsample of groups}, maximising the number of available satellites, and should minimise any scatter within the stacked mass profile (so long as the groups are indeed dominated by NFW halos). \autoref{eq:r rescale virial} and \autoref{eq:M rescale virial} therefore provide the optimum rescaling methodology to test for any deviations from the NFW profile. 

As a comparison, the more standard approach of rescaling solely by $M_{200}$ and $r_{200}$ produces mass profiles which still differ according to the halos' concentrations. In this case the NFW prediction and comparisons between different samples can become ambiguous. An obvious example would be comparing a low-mass and high-mass sample, where the mass-concentration relation would induce a split between the two enclosed mass profiles, even if the underlying physics are identical. For our purposes, the rescalings outlined by \autoref{eq:r rescale virial} and \autoref{eq:M rescale virial} provide a clearer test of NFW predictions.

We discuss the feasibility of measuring group concentrations in \autoref{subsubsec:discussion concentrations}. For now though, we will use exact concentrations for individual simulated groups which we calculate below in \autoref{subsubsec:concentration}.

\subsubsection{Calculating concentrations}\label{subsubsec:concentration}

Whilst there is a supplementary catalog which contains halo concentrations ($c_{200}$) for TNG-300 \citep{c200_calc_old}\footnote{See `(q) Halo Structure' at \url{https://www.tng-project.org/data/docs/specifications/}}, the concentrations provided are for the central subhalo. Ideally, we instead need the concentration of the FoF group halo for our analysis. Also, some groups have been removed from this catalog due to quality cuts. We would like to avoid such cuts at this stage as these might remove e.g. merging systems which could have an important impact on the measurement of dynamical mass profiles.

Therefore, we calculate the concentrations of groups ourselves using the same method as \citet{c200_calc}. To summarise, they separate the group's \textit{dark matter particles} into 50 spherical shells, equally spaced in logarithmic $r$. They chose the innermost edge as the larger of $0.01r_{200}$ or 3 times the gravitational softening length $\epsilon$ of the simulation ($\epsilon=1$ kpc/$h$ for TNG-300 at $z=0$), and the outermost edge as $1.2r_{200}$. The density of each shell can be calculated as
\begin{equation}
    \rho_{shell} = \frac{3M_{shell}}{4\pi \left(r_{outer}^{3} - r_{inner}^{3} \right)},
\end{equation}
where $M_{shell}$ is the mass of dark matter particles within the shell, and $r_{outer}$ and $r_{inner}$ are the outer and inner radii of the shell respectively. Poisson noise is assumed to dominate the uncertainty of the shell density such that
\begin{equation}
    \alpha_{\rho} = \frac{\rho_{shell}}{\sqrt{N_{shell}}},
\end{equation}
for $N_{shell}$ particles within the shell. They then fit these density profiles to an NFW profile (\autoref{eq:NFW rho}), with $\rho_{0}$ and $r_{s}$ as the free parameters, using a $\chi^{2}$ minimisation with the Levenberg-Marquardt algorithm to determine the best fitting values. They define the rms of a given fit as
\begin{equation}
    \rho_{rms} \equiv \sqrt{\frac{1}{n}\sum_{i}^{n} \left[ \ln (\rho_{i}) - \ln (\rho_{fit})\right]^{2}},
\end{equation}
where $n$ is the number of shells, $\rho_{i}$ is the density of the $i$th shell, and $\rho_{fit}$ is the corresponding density of the fitted NFW profile. The best fitting concentration can then be determined from the best fitting $r_{s}$ as $c_{200} = r_{200}/r_{s}$. We replicate this methodology from \citet{c200_calc} to calculate the concentrations of individual TNG-300 groups.

\subsubsection{Rescaled mass profiles}
\begin{figure}
	\includegraphics[width=\columnwidth]{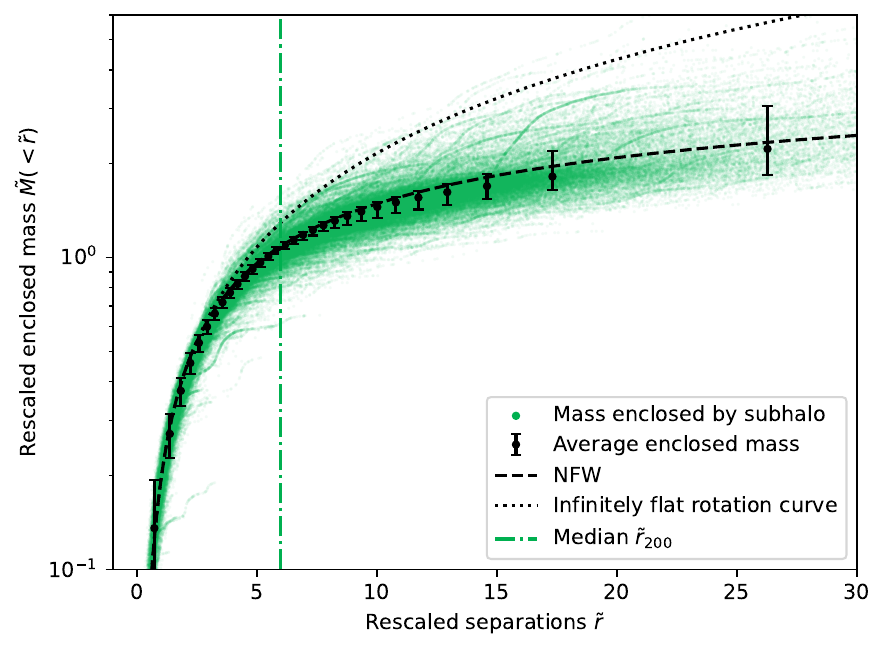}
    \caption{Rescaled true mass profiles for the 25,000 groups we analyse in TNG-300. The green points represent the rescaled mass contained below individual satellites (674341 in total). The black points with errorbars represent the median enclosed mass and the $1\sigma$ scatter about this median at a given rescaled separation. All particle types are used to calculate the enclosed mass. The dashed black line is the predicted rescaled NFW mass profile (\autoref{eq:Mass rescaled}), whilst the dotted black line represents an infinitely flat rotation curve model defined by \autoref{eq:Mass flat RC}. The green vertical dashdotted line represents the median host group virial radius in the rescaled space. The median true mass profile is within $1\sigma$ of the NFW profile at all radii and the scatter is small. The NFW and `infinitely flat rotation curve' profiles are also similar below the median $\tilde{r}_{200}$ and become increasingly divergent at larger $\tilde{r}$.}
    \label{fig:Rescale true profile}
\end{figure}

With the concentrations calculated in \autoref{subsubsec:concentration}, we can now test the rescalings defined by \autoref{eq:r rescale virial} and \autoref{eq:M rescale virial}. As a first step, we calculate and show the rescaled $\tilde{M}(<\tilde{r})$ profiles for the 25,000 analysed TNG-300 groups in \autoref{fig:Rescale true profile}, which demonstrates the effectiveness of this procedure. Each green point represents the rescaled mass enclosed below a subhalo, calculated as the summed mass of all particles within the group at a separation smaller than the subhalo's. The black points with errorbars show the resulting median mass profile and $1\sigma$ scatter about this median, within shells containing equal numbers of subhaloes. The median rescaled mass profile closely follows and is consistent with the NFW prediction described by \autoref{eq:Mass rescaled}, although the correspondence is not perfect as the median profile lies slightly below the NFW prediction. This mismatch arises due to the presence of the baryonic matter, which does not exactly follow the NFW profile. Still, the scatter around the median remains small ($\lesssim0.1$ dex to a few times the median $\tilde{r}_{200}$) and the NFW prediction does appear reasonable.

We also include a model for a system with an `infinitely flat rotation curve' as a reference. This model is identical to an NFW system (\autoref{eq:Mass rescaled}) up until some radius $\tilde{r}_{max}$, where the rotation curve reaches its maximum circular velocity $\tilde{v}_{max}$. The mass profile is then defined such that the rotation curve remains flat indefinitely beyond $\tilde{r}_{max}$. Mathematically, this is
\begin{equation}\label{eq:Mass flat RC}
    \tilde{M}_{flat}(<\tilde{r}) = \begin{cases}
        \ln(1+\tilde{r}) - \frac{\tilde{r}}{1+\tilde{r}}, & \tilde{r} \leq \tilde{r}_{max}, \\
        \tilde{v}_{max}^{2} \tilde{r}/\tilde{G}, & \tilde{r} > \tilde{r}_{max},
    \end{cases}
\end{equation}
(with $\tilde{G}=1$ by definition in the rescaled space). This model increasingly diverges from NFW and from the median mass profile as $\tilde{r}$ increases. Recovering accurate dynamical mass profiles out to as large $\tilde{r}$ as possible will therefore be important in decisively confirming or ruling out the infinitely flat rotation curves observed with weak lensing \citep{Brouwer_2021,Mistele_2024_RAR,Mistele_2024}.

\subsubsection{Velocities and line-of-sight data}\label{subsubsec:pos and vel data}

The final aspect of rescaling the data to discuss is the measurement of velocities and line-of-sight data to which \autoref{eq:Jeans 1-D} can be applied.

For velocities, we calculate the relative 3-D velocities between subhalos and their central with the Hubble flow contribution included. For a given vector separation $\mathbf{r}$ and vector peculiar velocity $\mathbf{v}_{pec}$ between the subhalo and central, the physical velocity of the subhalo will be
\begin{equation}
    \mathbf{v} = \mathbf{v}_{pec} + H_{0} \mathbf{r}.
\end{equation}
We measure line-of-sight velocities by setting the x-axis as the line-of-sight for simplicity. These will also include contributions from the Hubble flow, so are calculated as
\begin{equation}\label{eq:v los}
    v_{los} = v_{x} + H_{0} x,
\end{equation}
where $x$ and $v_{x}$ are the physical separation and relative peculiar velocity in the x-direction respectively. We set an uncertainty of $u_{v} = 10$ km/s on all line-of-sight velocities as this is the expected uncertainty for a modern survey like DESI BGS \citep{DESI_validation_2023}. The uncertainty on the relative line-of-sight velocity is the sum in quadrature of the group and subhalo uncertainties, so is
\begin{equation}\label{eq:v uncertainty}
    u_{vlos} = \sqrt{2u_{v}^{2}}.
\end{equation}
Projected separations are calculated as 
\begin{equation}\label{eq:R projected}
    R = \sqrt{y^{2} + z^2},
\end{equation}
where $y$ and $z$ are the separation between the central and subhalo position in the y and z directions respectively.

The rescaled projected separations are calculated according to \autoref{eq:r rescale virial}, whilst the rescaled line-of-sight velocities and corresponding uncertainties are calculated through \autoref{eq:v rescale} and \autoref{eq:v unc rescale} as
\begin{equation}
    \tilde{v}_{los} = v_{los}\sqrt{\frac{r_{s}}{GM_{s}}},
\end{equation}
and 
\begin{equation}
    \tilde{u}_{vlos} = \sqrt{2u_{v}^{2}\frac{r_{s}}{GM_{s}}}.
\end{equation}
We calculate the 3-D physical velocities, line-of-sight velocities and projected separations for all subhalos within our 25,000 groups.

\subsection{Quality cuts}\label{subsec: Quality cuts}

Before applying the Jeans equation to the TNG-300 subhalos, we must perform some quality cuts to the data. Broadly speaking, we first attempt to remove subhalos which may have unreliable properties within the group catalogue. For example, low mass subhalos near the resolution limit may not constitute bound systems between simulation snapshots. From the remaining subhalos, we then compare some simple, physically motivated cuts which attempt to remove systematic effects from the dynamical data.

To remove unreliable subhalos, we cut based on the subhalo mass. Higher mass subhalos should be more strongly gravitationally bound and so are more likely to remain bound between snapshots. They should also possess more reliable velocity measurements as subhalos velocities are calculated as the average of the member particles' velocities within the TNG-300 group catalogue. More member particles should equate to a more reliable velocity. We try two separate such cuts.

For the first, we select subhalos with a \textit{total} mass greater than 100 times the dark matter mass resolution ($M\geq 4.0 \times 10^{9}M_{\odot}/h$). These subhalos will therefore contain at least 100 particles. We refer to the subsample this creates with the prefix `Total-' (for total mass cut). We find that 215867 subhalos ($\sim32\%$) pass this total mass cut.

For the second, we select subhalos with a \textit{stellar} mass greater than 100 times the baryonic mass resolution ($M_{*}\geq7.6 \times 10^{8} M_{\odot}/h$). These should correspond more closely to genuine satellite galaxies that could be observed (the total mass cut before would include many dark matter only systems). We will therefore refer to these structures as `satellites' and use the prefix `Sat-' to refer to the subsample this produces. We find that 34221 subhalos ($\sim5\%$) pass this stellar mass cut.

After these mass cuts, we consider further quality cuts that aim to remove dynamical systematics. The spherical Jeans equation assumes that the system is spherically symmetric and that the system is in dynamical equilibrium. Since observed groups are randomly oriented, a stacked sample should satisfy the assumption of spherical symmetry. The assumption of equilibrium is less clearly followed. Groups will become dynamically disturbed during merger events and are expected to accrete new matter over time. We attempt to address these issues by comparing some simple, physically motivated quality cuts at the group and subhalo levels respectively. 

In \autoref{subsubsec:Virialisation idenification} we perform group level cuts by identifying dynamically relaxed groups. In \autoref{subsubsec:Infall time data} and \autoref{subsubsec: quenched satellites}, we perform subhalo level cuts by identifying subhalos that infell onto their host group long ago and those that are quenched respectively. An earlier infall time provides more time for a subhalo to have phase-mixed (i.e. reached dynamical equilibrium) with its group. Satellites are also expected to quench a few Gyr after infall \citep{Hough_2023}, so quenched satellites should provide an observationally accessible proxy for those that infell long ago.

In total we will have seven samples to compare: three when performing a total mass cut on subhalos and four when performing a stellar mass cut to identify satellites. We call these `Total-All' when only the total mass cut is performed, `Total-Virial' when the total mass cut is performed on subhalos from dynamically relaxed groups, and `Total-Infall' when the total mass cut is performed on subhalos which infell long ago. Similarly, when performing the stellar mass cut we have `Sat-All', `Sat-Virial', and`Sat-Infall', along with a `Sat-Quench' sample when selecting for quenched satellites. We additionally apply an interloper removal algorithm (\autoref{subsec:Interlopers}) and a radial cut (\autoref{subsec:Radial cut}) to each of these seven samples to remove any remaining outliers. We will summarise this information again at the end of the section in \autoref{subsubsec:Cut samples}.

\subsubsection{Dynamically relaxed groups}\label{subsubsec:Virialisation idenification}

The spherical Jeans equation assumes that the system is in dynamical equilibrium (or `relaxed'). To identify such groups, we use the three criteria defined by \citet{Neto_2007}.

The first criterion is for the virialisation of the group and is the most involved to calculate. For a dynamically relaxed system with mass $M$, the scalar virial theorem predicts that the total kinetic energy $T$ and the total potential energy $U$ are related by \citep{Davis_2011}
\begin{equation}\label{eq:Scalar virial theorem}
    0 = \left(2T - S_{p}M\right) + \left(U + U_{ext} \right),
\end{equation}
with $U_{ext}$ and $S_{p}$ accounting for an external gravitational potential and surface pressure respectively. Surface pressure in this context corrects for the fact that the virial radius is not a physical boundary and particles are free to cross it, so not all particles considered are bound by the potential at the virial radius. $U_{ext}$ is found to have negligible contribution for dark matter halos \citep{Davis_2011}, so can be neglected. A `virial ratio' $\eta$ can therefore be defined as
\begin{equation}\label{eq:virial ratio}
    \eta \equiv - \frac{2T - S_{p}}{U}.
\end{equation}
Systems with $\eta=1$ would be perfectly virialised, those with $\eta < 1$ would have too little kinetic energy to be virialised (so collapse) and those with $\eta > 1$ would have too much kinetic energy to be virialised (so expand). \citet{Neto_2007} enforce a cut of $\eta <1.35$ to identify virialised groups. To determine whether a group meets this criterion, we calculate the virial ratio using all member particles within $r_{200}$. We implicitly assume in this step that virialisation of the simulation particles corresponds to virialisation of the subhalos.

The total kinetic energy is trivial to measure as
\begin{equation}
    T = \frac{1}{2}\sum_{i} m_{i}v_{i}^{2},
\end{equation}
where $m_{i}$ and $v_{i}$ are the mass and speed of the $i$th particle respectively, and the sum is over all particles within $r_{200}$.

The potential energy is more computationally expensive to calculate directly. The exact value is a sum over all particle pairs as
\begin{equation}\label{eq:Potential energy pairs}
    U = \sum_{i=1}^{N-1}\sum_{j=i+1}^{N} - G \frac{m_{i}m_{j}}{|\mathbf{r}_{i}-\mathbf{r}_{j}|},
\end{equation}
where $m_{i}$ are the particle masses and $\mathbf{r}_{i}$ are their vector positions. This scales as $\mathcal{O}(N^{2})$ and will become impractical to compute for massive halos containing many particles. Instead, we follow \citet{Neto_2007} and calculate the gravitational potential energy between 1000 randomly selected particles. For $n$ randomly selected particles in a system of $N$ total particles, the total gravitational potential energy can be estimated from this subset by
\begin{equation}
    U = \frac{N(N-1)}{n(n-1)}\sum_{i=1}^{n-1}\sum_{j=i+1}^{n} - G \frac{m_{i}m_{j}}{|\mathbf{r}_{i}-\mathbf{r}_{j}|}.
\end{equation}
For each group, we randomly select 1000 particles within $R_{200}$ (i.e. $n=1000$) and set $N$ equal to the total number of particles within $r_{200}$ to estimate its total potential energy.

The surface pressure term is given by \citet{Klypin_2016} as
\begin{equation}\label{eq:surface pressure}
    S_{p} = \frac{1}{M_{200}} \oint \rho \mathbf{x} \cdot \mathbf{vv} \cdot d\mathbf{S} = \frac{4\pi r_{200}^{3}\rho(r_{200})\sigma_{r}^{2}(r_{200})}{M_{200}},
\end{equation}
when taken over the surface of a sphere with radius $r_{200}$. The density and radial velocity dispersion at $r_{200}$ can be calculated by binning the data. We use the same spherical shells as defined when calculating the concentration in \autoref{subsubsec:concentration} to construct density and radial velocity profiles of the simulation particles. We then linearly interpolate these profiles to $r_{200}$ to estimate $\rho(r_{200})$ and $\sigma_{r}^{2}(r_{200})$. The inclusion of the surface pressure term causes a downwards shift in the virial ratio, compared to using only the kinetic and potential energy terms. This corrects a rise in virial ratios at increasing halo mass (see Figure 2 of \citeauthor{Neto_2007} \citeyear{Neto_2007} and Figure 5 of \citeauthor{Klypin_2016} \citeyear{Klypin_2016}) that can lead to a preferential misidentification of the virialisation of massive systems.

\begin{figure*}
\includegraphics[width=\textwidth]{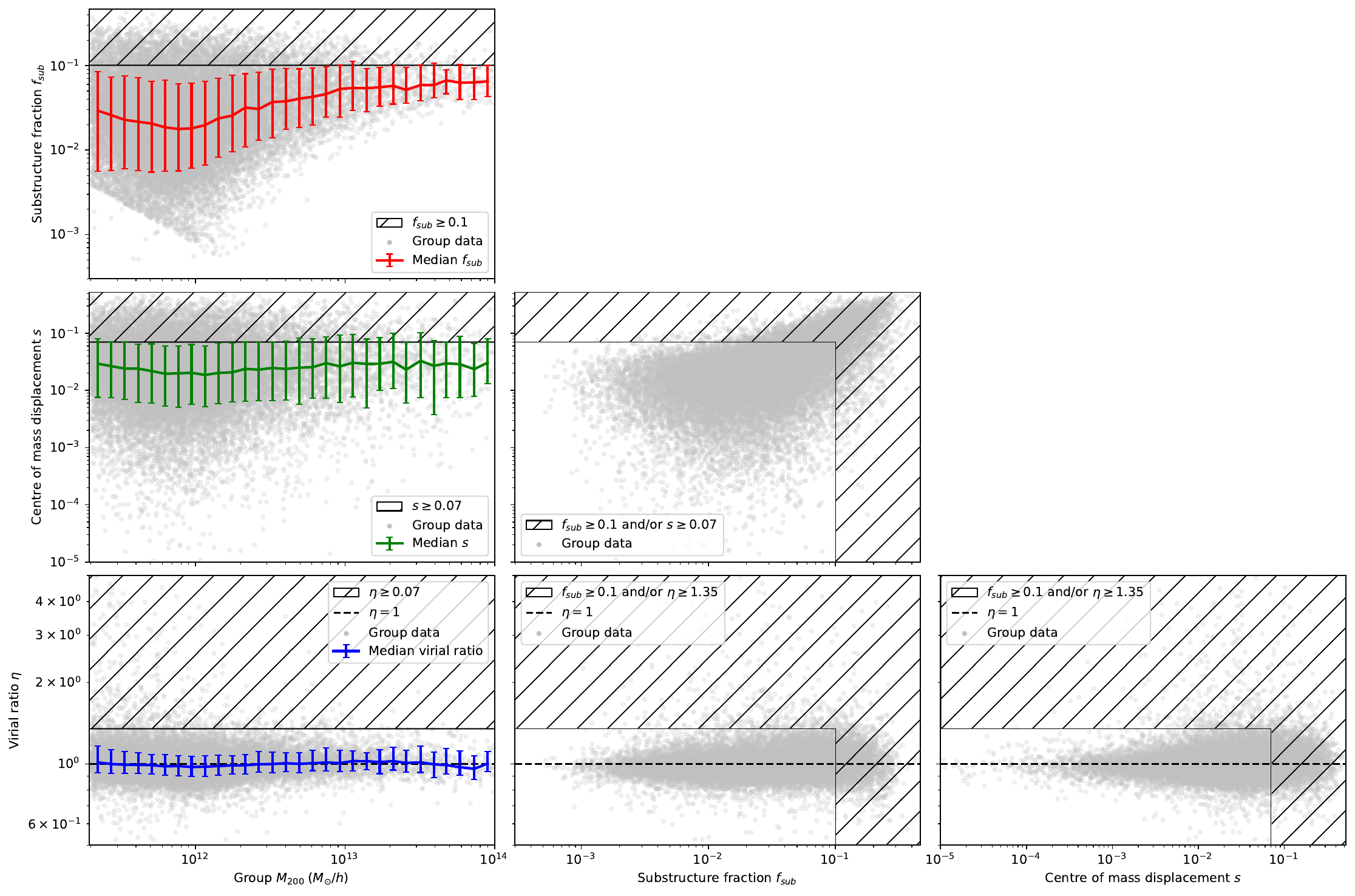}
    \caption{Scatter plots showing the relations of virial ratios $\eta$, centre of mass displacements $s$, and fraction of mass in substructure $f_{sub}$ of TNG-300 groups at $z=0$ with respect to group $M_{200}$ and each other. Grey points indicate the values of individual groups whilst hatched regions denote where groups are defined as dynamically unrelaxed (see \autoref{subsubsec:Virialisation idenification}). Distributions with respect to $M_{200}$ are shown in the leftmost column. The coloured lines with errorbars show median values and the $1\sigma$ scatter about these medians at increasing $M_{200}$. In the top panel, the median $f_{sub}$ increases with mass whilst the scatter decreases. In the middle panel, $s$ shows little dependence on $M_{200}$ and the same is true for $\eta$ in the bottom panel. The median value of $\eta$ is close to perfect virialisation ($\eta=1$) at all masses. The middle column shows the relation between $f_{sub}$ with $s$ (middle row) and $\eta$ (bottom row) respectively. There is a visible positive correlation between $f_{sub}$ and $s$ but little visible correlation between $f_{sub}$ and $\eta$. The right column shows the relation between $s$ and $\eta$, with negligible visible correlation between them.}
    \label{fig:Virial hist}
\end{figure*}

The second criterion outlined by \citet{Neto_2007} is the `centre of mass displacement'. This is the offset between the position of the minimum potential $\mathbf{r}_{c}$ (i.e. the halo position in the TNG-300 group catalogues) and the centre of mass $\mathbf{r}_{cm}$ of all particles within $R_{200}$. This is characterised by \citep{Thomas_2001}
\begin{equation}
    s = \frac{|\mathbf{r}_{c} - \mathbf{r}_{cm}|}{R_{200}},
\end{equation}
with a cut of $s<0.07$ for relaxed groups. The logic here is that the position of minimum potential should remain stable during merger or flyby events, whilst the centre of mass will be displaced at the start of such events by the mass of the merging/flyby halo. Such merger/flyby events can drive a relaxed halo out of dynamical equilibrium.

The third and final criterion of \citet{Neto_2007} is the fraction of mass in substructure $f_{sub}$. This is the total mass of subhalos found within $R_{200}$ (excluding the central) divided by the $M_{200}$ of the group, with a cut of $f_{sub} < 0.1$ for relaxed groups. This substructure fraction should be expected to decrease over time for a stable system as dynamical friction causes substructure to infall and coalesce with the primary halo. A large substructure fraction is therefore indicative of a recent merger event.

We show the relations of $\eta$, $s$ and, $f_{sub}$ with group $M_{200}$ and with each other in \autoref{fig:Virial hist}. To highlight the interesting relations, we can first see that $s$ and $\eta$ appear to have negligible dependence on the group $M_{200}$ and that the median $\eta$ remains close to perfect virialisation ($\eta=1$) at all masses. Meanwhile, the median $f_{sub}$ shows an increase with mass, but the scatter about this median also decreases with mass. The largest values of $f_{sub}$ are therefore actually found within low mass halos. Visually, there is a sharp diagonal edge in the bottom left of this $f_{sub}$ distribution. This edge denotes groups containing one subhalo (in addition to their central) at the resolution limit since subhalos must contain at least 20 particles to be included in the TNG-300 catalog \footnote{See \url{https://www.tng-project.org/data/docs/faq/}}, with a few groups lying below this edge since star/gas particles can be lighter than dark matter particles. The disappearance of this edge at $M_{200}\approx 10^{12}M{\odot}/h$ indicates that groups more massive than this all contain significant substructure.

There is a clear positive correlation between $s$ and $f_{sub}$ in \autoref{fig:Virial hist}. This is expected if both are indicators of merger/flyby events. There also appears to be some correlation of the virial ratio with $s$ and $f_{sub}$, whereby a larger scatter in the values of $\eta$ is found at larger values of $s$ and $f_{sub}$. Again this would be expected if large $s$ and $f_{sub}$ arise when the group is disrupted in some way. These results are broadly consistent with Figure 2 of \citet{Neto_2007} once the surface pressure correction to the virial ratios we include here are taken into account (\autoref{eq:virial ratio} and \autoref{eq:surface pressure}).

To summarise, a group is considered dynamically relaxed if $\eta<1.35$, $s<0.07$ and $f_{sub} < 0.1$. With these cuts, we find that 19,893 of the 25,000 ($\sim 80\%$) TNG-300 groups we selected in \autoref{subsec:TNG groups} are dynamically relaxed. In terms of subhalos, we find that $\sim 73\%$ of subhalos which pass our total mass cut are found within relaxed groups (157235) and that $\sim 68\%$ of satellites which pass our stellar mass cut are found within relaxed groups (23178).

\subsubsection{Infall time of subhalos}\label{subsubsec:Infall time data}

A recently accreted subhalo will not be phase mixed with its group. This is only achieved after many gravitational interactions between the subhalo and virialised group matter, either through dynamical friction or 2-body interactions with other substructure. The timescales for such processes are usually given as a factor of the crossing time $t_{cross}$, defined by \citet{Binney_2008} as
\begin{equation}
    t_{cross} \equiv \left(\frac{R_{200}^{3}}{GM_{200}} \right)^{1/2} = \left(\frac{3}{800\pi G \rho_{c}} \right)^{1/2}.
\end{equation}
This is the time it would take a particle with constant speed equal to the circular velocity at $R_{200}$ to cross a distance equal to $R_{200}$. Since by definition $M_{200} \propto R_{200}^{3}$, the crossing time for all TNG-300 groups at $z=0$ is $\approx 1.4$ Gyr.

We define the infall time of a subhalo onto a group as the first simulation snapshot in which the subhalo and the $z=0$ central are in the same group. We adopt $t_{infall} > 3 t_{cross} \approx 4.3$ Gyr for our quality cut. This represents the approximate limit where further cuts provide diminishing changes to the recovered mass profile when applying the 3-D Jeans equation (\autoref{eq:Jeans 3D}), whilst also retaining many subhalos to actually analyse. We find $\sim 41\%$ of subhalos which pass our total mass cut (89276) infell more than $3t_{cross}$ ago, whilst $\sim 53\%$ of satellites which pass our stellar mass cut (18243) infell more than $3t_{cross}$ ago.

\subsubsection{Quenched satellites}\label{subsubsec: quenched satellites}
Satellites are expected to quench a few Gyr after accretion onto a group \citep{Hough_2023}, so quenched satellites should provide a rough (and more realistically attainable) proxy of satellites which infell long ago. We define a satellite as quenched when its specific star formation rate (i.e. star formation rate divided by total stellar mass) is less than $10^{-11}$ yr$^{-1}$. We only select from satellites that passed our stellar mass cut and find that $\sim 46\%$ of these (15781) are quenched.

\subsection{Removing outliers}

After producing the seven distinct samples above (\autoref{subsec: Quality cuts}), we impose additional checks to remove any obvious remaining outliers. The virial shape parameters described in \autoref{eq:4th order beta 1} and \autoref{eq:4th order beta 2} are dependent on the fourth line-of-sight velocity moment $\langle v_{los}^{4} \rangle$ and such higher order moments are susceptible to biasing from even a small number of outliers. Therefore, we perform an interloper analysis in \autoref{subsec:Interlopers} followed by a radial cut in \autoref{subsec:Radial cut}.

\subsubsection{Interloper removal}\label{subsec:Interlopers}

\begin{figure*}
	\includegraphics[width=\textwidth]{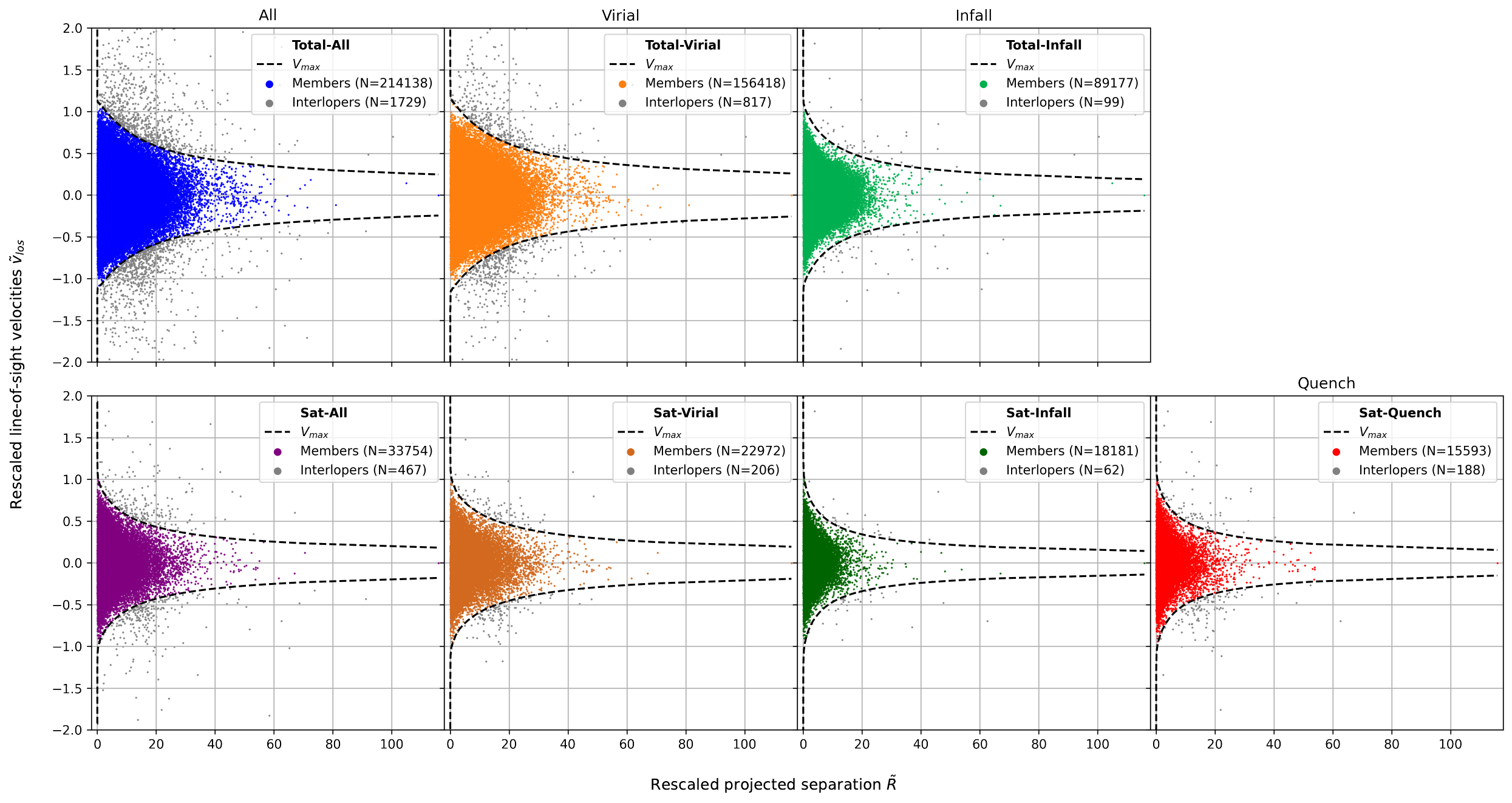}
    \caption{Phase-space diagrams of subhalos corresponding to the samples defined in \autoref{subsec: Quality cuts}. The top row shows the \textit{Total-} samples and the bottom row shows the \textit{Sat-} samples. From left to right the columns show the \textit{-All}, \textit{-Virial}, \textit{-Infall}, and \textit{-Quench} samples. Sample names are also provided in the legend of each panel. The layout and colours used here will be replicated in later figures where appropriate. The individual points in each panel show the rescaled line-of-sight velocities and projected separations of subhalos within that sample. The black dashed lines show the converged iterations of the $V_{max}$ profiles (\autoref{subsec:Interlopers}) for each sample. Coloured points between these $V_{max}$ profiles are considered `member' subhalos, whilst the grey points outside these profiles are considered `interlopers' and removed from subsequent analyses. The number of member and interloper subhalos for each sample is provided in the legends. The number of interlopers ranges from $0.1\%$ to $1.4\%$ of a given sample.}
    \label{fig:Interlopers}
\end{figure*}

For galaxy groups and clusters, satellite galaxies which are not gravitationally bound to the system are known as interlopers and will appear as outliers compared to the bound population. There are various methods for identifying interlopers \citep[see][]{Wotjak_2007} and here we adopt a less prohibitive implementation of the method outlined by \citet{Hartog_1996}, which compares the maximum velocity that gravitationally bound satellites could possess to their observed line-of-sight velocity.

To estimate this maximum velocity, \citet{Hartog_1996} first consider a stable system where all tracers travel on circular orbits. In this case, the virial theorem applies such that $2T + U=0$. All tracers will travel at the circular velocity
\begin{equation}\label{eq:vcirc}
    v_{c}^{2}(r) = \frac{GM(<r)}{r},
\end{equation}
in this case. Next, they consider a tracer on an infalling trajectory, which to be gravitationally bound must satisfy the condition $U_{i}+T_{i} < 0$. Comparing these two cases, the maximum infall velocity of bound particles must therefore be $v_{inf}(r) = \sqrt{2}v_{c}(r)$. When projecting along the line-of-sight, a tracer's physical separation could be positioned at any $r>R$. \citet{Hartog_1996} therefore define an $r_{max}$ as the physical separation along the line-of-sight at which $v_{inf}$ is maximised and so defined the maximum bound line-of-sight velocity as
\begin{equation}\label{eq: vmax}
    V_{max}(R) \equiv v_{inf}(r_{max}).
\end{equation}
For a given $M(<r)$ profile, a tracer with $v_{los} > V_{max}(R)$ can be identified as unbound.

However, a crude measure of $M(<r)$ is required for this test, for which \citet{Hartog_1996} use a virial theorem mass estimator of the form \citep{Bachall_1981}
\begin{equation}\label{eq:MVT}
    M_{VT}(<R) = \frac{3\pi N(<R)}{2G} \frac{\sum_{i}v_{i}^{2}}{\sum_{i<j}1/R_{ij}}.
\end{equation}
$N$ is the number of tracers and the $R_{ij}$ term is the projected separation between the $i$th and $j$th tracers. A mass profile can be constructed by calculating $M_{VT}$ for all tracers within increasingly larger $R$, although this formula will overpredict the mass profile even for pristine tracer samples \citep[due to the lack of surface pressure term;][]{Biviano_2006,Wotjak_2007}. Since it is difficult to perfectly identify all interlopers, this is considered acceptable as an overestimate in $V_{max}$ can still remove most interlopers (and reduce the bias in the velocity moments) without removing genuine bound tracers.

The general interloper removal procedure is therefore to calculate $M_{VT}(<R)$ at the projected separation of each tracer using \autoref{eq:MVT} so that the $V_{max}(R)$ for each tracer can be estimated with \autoref{eq: vmax}. All tracers with $|v_{los}| > V_{max}(R)$ are then removed from the analysis. This process is repeated iteratively until convergence.

We apply this procedure to the samples described in \autoref{subsec: Quality cuts} to identify any remaining interlopers. We stack each sample and use the rescaled line-of-sight velocities $\tilde{v}_{los}$ and rescaled projected separations $\tilde{R}$ to calculate rescaled $\tilde{M}_{VT}$ and $\tilde{V}_{max}$ profiles. When calculating the $\tilde{R}_{ij}$ terms, we include a softening length $\epsilon$ as $\sum_{i<j}1/R_{ij}$ can become spuriously dominated by individual tracers which lay very close in projected space. We define this softening length as $1\%$ of the mean interparticle distance such that
\begin{equation}
    \epsilon \equiv 0.01 \sqrt{\frac{\pi \tilde{R}_{end}^{2}}{N}},
\end{equation}
where $\tilde{R}_{end}$ is the maximum rescaled projected separation of any tracer. We therefore calculate
\begin{equation}
    \tilde{R}_{ij} = \sqrt{(\tilde{y}_{i} - \tilde{y}_{j})^{2} + (\tilde{z}_{i} - \tilde{z}_{j})^{2} + \epsilon^{2}}.
\end{equation}
\citet{Hartog_1996} also provide more restrictive criterion for $V_{max}$ (i.e smaller $V_{max}$) past the projected virial radius. We find these remove many bound satellites when applied to our stacked samples, so we use only the less stringent \autoref{eq: vmax}, regardless of projected separation.

We show the results of this interloper removal for each sample in \autoref{fig:Interlopers}. As seen in the legends, the number of interlopers identified in each sample is low, ranging from $0.1\%$ to $1.4\%$ of a given sample. We can also start to see the effects of each quality cut. Selecting satellites which infell long ago leads to a smaller fraction of interlopers compared to selecting just for relaxed groups, which itself produces a lower fraction of interlopers compared to no quality cuts.

\subsubsection{Radial cut}\label{subsec:Radial cut}

After the interloper removal in \autoref{subsec:Interlopers}, we finally apply a radial quality cut to each sample. Subhalos at the largest $\tilde{R}$ will either originate from the host groups with the largest concentrations or represent objects with physical separations several times the host group virial radius (so unlikely to be phase-mixed). In either case these are likely undesirable outliers from the general population. We therefore remove any subhalos at a $\tilde{R}$ greater than three times the median host group $\tilde{r}_{200}$ of a given sample.

The inverse is less problematic. Subhalos at low $\tilde{R}$ are either likely to be within their host's virial radius (so likely to be phase-mixed) or originate from the host groups with the lowest concentrations. The general population should dominate the dynamics at low $\tilde{R}$ over any groups with exceptionally low concentrations.

We summarise the effects of this radial cut in \autoref{tab:radial cut}, where we also show the median host $\tilde{r}_{200}$ for each sample. The number of subhalos removed by this cut is between $0.7\%$ and $4.1\%$ per sample, whilst the median host $\tilde{r}_{200}$ shows some variance between different samples. 

A lower median $\tilde{r}_{200}$ shows a sample preferentially selects subhalos from lower concentration groups. This is a direct result of the rescaling defined in \autoref{eq:r rescale virial}, with the variance in median $\tilde{r}_{200}$ being driven by the relation between concentration and mass assembly history of the halo. Halos which assembled their mass late (i.e. through recent mergers) will have lower concentrations than those which assembled their mass early \citep{Wechsler_2002}. This produces the mass-concentration relation, where higher mass halos tend to have lower concentrations \citep[this is visible for TNG300 in Figure 3 of][]{c200_calc_old}, and a relation where dynamically relaxed halos tend to have higher concentrations \citep[see Figure 6 of][]{Neto_2007}. 

The host mass distribution of each sample is shown in \autoref{fig:Group mass hist}, where we can see that the \textit{Sat-} samples are biased to higher masses than their \textit{All-} counterparts, as substructures in low mass groups will themselves be low in mass so are more likely to be removed by the more restrictive stellar mass cut. Each \textit{Sat-} sample therefore has a comparatively lower median $\tilde{r}_{200}$ (\autoref{tab:radial cut}) as a result of this mass bias, whilst the \textit{Sat-Quench} sample has the strongest bias to high mass and the lowest median $\tilde{r}_{200}$ overall. Similarly, each \textit{-Virial} and \textit{-Infall} sample has a comparatively higher median $\tilde{r}_{200}$ than its \textit{-All} counterpart due to the explicit selection for dynamically relaxed systems.

\begin{table}
    \centering
    \caption{Median host $\tilde{r}_{200}$ and number of subhalos removed by radial cut for each sample (after interloper removal has been performed).}
    \label{tab:radial cut}
    \begin{tabular}{lcc}
        \hline
        Sample & Median host $\tilde{r}_{200}$ & Removed by radial cut \\
        \hline
        \multicolumn{3}{l}{Total mass cut} \\
        \textit{Total-All} & 5.92 & 6159 $(2.9\%)$ \\
        \textit{Total-Virial} & 6.40 & 4128 $(2.6\%)$ \\
        \textit{Total-Infall} & 6.00 & 778 $(0.9\%)$ \\
        \\[0.5em]
        \multicolumn{3}{l}{Stellar mass cut} \\
        \textit{Sat-All} & 5.54 & 1318 $(3.9\%)$ \\
        \textit{Sat-Virial} & 6.00 & 937 $(4.1\%)$ \\
        \textit{Sat-Infall} & 5.62 & 123 $(0.7\%)$ \\
        \textit{Sat-Quench} & 5.36 & 362 $(2.3\%)$ \\
        \hline
    \end{tabular}
\end{table}

\subsection{Final subhalo samples}\label{subsubsec:Cut samples}

After quality cuts we are left with seven distinct subhalo samples, from each of which we randomly select 10,000 subhalos for a fairer comparison in the dynamical analysis. This is approximately an order of magnitude lower than the number of tracers that would be available after reasonable quality cuts with a modern redshift survey like DESI BGS, but we are limited in this case by the available simulation data (\textit{Sat-Quench} contains only $\sim 15,000$ subhalos after all cuts are performed). Dynamical analyses using 10,000 tracers are still sufficient for this work, as we are more interested in first testing the viability of the stacking and rescaling method.

\begin{figure}
\includegraphics[width=\columnwidth]{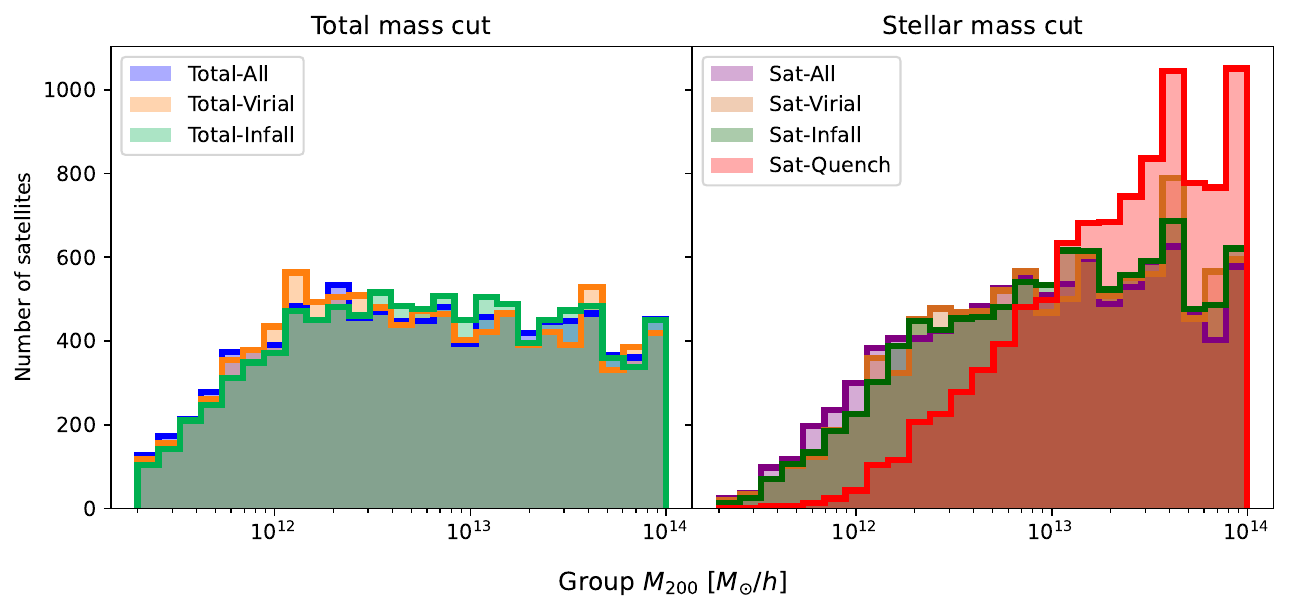}
    \caption{Histograms showing the host group masses ($M_{200}$) for the 10,000 satellites selected in each of our samples. The left panel shows the samples where a total mass cut is performed. The \textit{Total-All}, \textit{Total-Virial}, and \textit{Total-Infall} samples show near identical host mass distributions. The right panel shows the samples where a stellar mass cut is performed. The \textit{Sat-All}, \textit{Sat-Virial}, and \textit{Sat-Infall} samples show near identical distributions to each other, but preferentially select subhalos in higher mass groups compared to their \textit{Total-} counterparts. The \textit{Sat-Quench} sample shows an even stronger skew to subhalos in high mass groups.}
    \label{fig:Group mass hist}
\end{figure}

In \autoref{fig:Group mass hist} we show the distribution of host group masses for the 10,000 randomly selected subhalos in each sample. The `\textit{-Virial}' and `\textit{-Infall}' samples do not alter the host group mass distribution when compared to the `\textit{All}' samples, whilst the `\textit{-Quench}' sample preferentially selects subhalos from higher mass groups. The `\textit{Sat-}' samples also preferentially select subhalos from higher mass groups compared to their `\textit{Total-}' counterparts.

\section{Methods}\label{sec:Methods}
We use two publicly available codes for the dynamical analysis, called \textsc{binulator} \citep{Gravsphere_binulator} and \textsc{gravsphere} \citep{Gravsphere_1, Gravpshere_2, Gravsphere_3}\footnote{\url{https://github.com/justinread/gravsphere}}. \textsc{binulator} measures $\sigma_{los}^{2}(R)$, $\Sigma(R)$, and $\langle v_{los}^{4} \rangle(R)$ from line-of-sight data in bins of increasing projected separation. \textsc{gravsphere} then uses these quantities to measure the dynamical mass profile through the line-of-sight Jeans equation. 

Both codes use Markov Chain Monte Carlo (MCMC) to fit model parameters to the data, with posteriors sampled using the affine invariant sampler \textsc{emcee} \citep{emcee_2013}. Errors are assumed to be Gaussian and uncorrelated in both codes, such that the likelihood is related to the $\chi^{2}$ distributions by $\mathcal{L}=\exp\left(-\chi^{2}/2 \right)$. The best fitting model parameters are found at maximum $\mathcal{L}$ (minimum $\chi^{2}$) and confidence intervals are determined directly from percentiles of the posterior distributions. 

We will detail the burn-in, number of walkers and number of models per walker throughout \autoref{subsec:Binulator} and \autoref{subsec:Gravsphere} when \textsc{emcee} is employed. We will also detail priors when relevant and summarise this information in \autoref{subsec:parameters summary}.

\subsection{Binulator}\label{subsec:Binulator}
\subsubsection{Binning line-of-sight data}
We measure the rescaled $\tilde{\Sigma}(\tilde{R})$, $\tilde{\sigma}_{los}^{2}(\tilde{R})$ and $\vzfour(\tilde{R})$ using the \textsc{binulator} code \citep{Gravsphere_binulator}, which is included with the \textsc{gravsphere} installation. 

For each subhalo we provide \textsc{binulator} with a rescaled projected separation $\tilde{R}$, a rescaled line of sight velocity $\tilde{v}_{los}$, an uncertainty on this velocity $\alpha_{\tilde{v}}$ and a probability weight $w_{i}$. We calculate $\tilde{R}$, $\tilde{v}_{los}$ and $\alpha_{\tilde{v}}$ as outlined in \autoref{subsec:Rescaling}. $w$ represents the probability that a subhalo belongs to the host group, so the effective number of tracers is defined as $\sum w$. For simplicity, we set $w=1$ for all subhalos.

Once line of sight data are provided, binulator separates the data into `photometric' and `kinematic' shells of increasing $R$. The `photometric' bins are used to measure surface density, whilst the `kinematic' bins are used to measure velocity dispersions and fourth-order velocity moments. Since all of our subhalos have velocity information, we set equal `photometric' and `kinematic' bin sizes of $N_{phot}=N_{kin}=100$ for the effective number of tracers per bin. Therefore we have 10,000 subhalos distributed into 100 bins. 

The bin edges are calculated by taking midpoint between the outermost subhalo in the bin and the innermost subhalo in the next bin, whilst the bin centres are calculated as the midpoint between the inner and outer edge of the bin. The inner edge of the first bin is set as $R=0$ and the outer edge of the final bin is set as the projected separation of the outermost subhalo.

\subsubsection{Surface density}\label{subsubsec:Surface density}

To model the surface density profiles, \textsc{binulator} measures the surface density of tracers within each radial bin and then fits an analytic profile. This fitting is performed using \textsc{emcee} and the default implementation uses a Plummer sphere model \citep{Plummer_1911},
\begin{equation}\label{eq:Plummer SD}
\Sigma_{p}(R)=\sum_{s}\frac{N_{s}a_{s}^{2}}{\pi\left(a_{s}^{2}+R^{2} \right)^{2}}.
\end{equation}
Here, $s$ represents a number of Plummer spheres to sum over, $N_{s}$ are the `masses' of each sphere, and $a_{s}$ are the radii of each sphere. This model is flexible as more spheres can be added to fit complex data. It also has a corresponding analytic number density profile (see \autoref{subsec:Number density}). 

The observed surface density of the $i$th bin is calculated as
\begin{equation}\label{eq:Observed SD}
    \Sigma_{obs,i} = \frac{N_{i}}{\pi\left(R_{out,i}^{2} - R_{in,i}^{2} \right)},
\end{equation}
where $N_{i}$ is the number of tracers in the bin and the inner and outer bin edges are $R_{in,i}$ and $R_{out,i}$ respectively. Poisson uncertainties are assumed such that
\begin{equation}
    \alpha_{\Sigma,i} = \frac{\Sigma_{i}}{\sqrt{N_{i}}}.
\end{equation}
The log-likelihood provided to \textsc{emcee} then takes the form
\begin{equation}\label{eq:Likelihood SD}
    \ln\left(\mathcal{L}_{\Sigma} \right)= -\frac{1}{2}\sum_{i} \frac{\left[\Sigma_{p}(R_{i}) - \Sigma_{obs,i} \right]^{2}}{\alpha_{\Sigma,i}^{2}}.
\end{equation}

We found three Plummer spheres (unchanged from the standard installation of \textsc{binulator}) is sufficient to fit our data. We therefore fit six surface density parameters, with priors set as: $10^{-4}<N_{1}<10^{2}$, $10^{-4}<N_{2}<10^{2}$, $10^{-4}<N_{3}<10^{2}$, $0.25<a_{1}<3$, $0.25<a_{2}<3$, and $0.25<a_{3}<10$. \textsc{binulator} priors can be viewed in \autoref{tab:Binulator}. We used 500 walkers, 10,000 models per walker and a burn-in of 75\% (unchanged from the standard installation of \textsc{binulator}) to fit the surface density profiles.

\begin{figure*}
	\includegraphics[width=\textwidth]{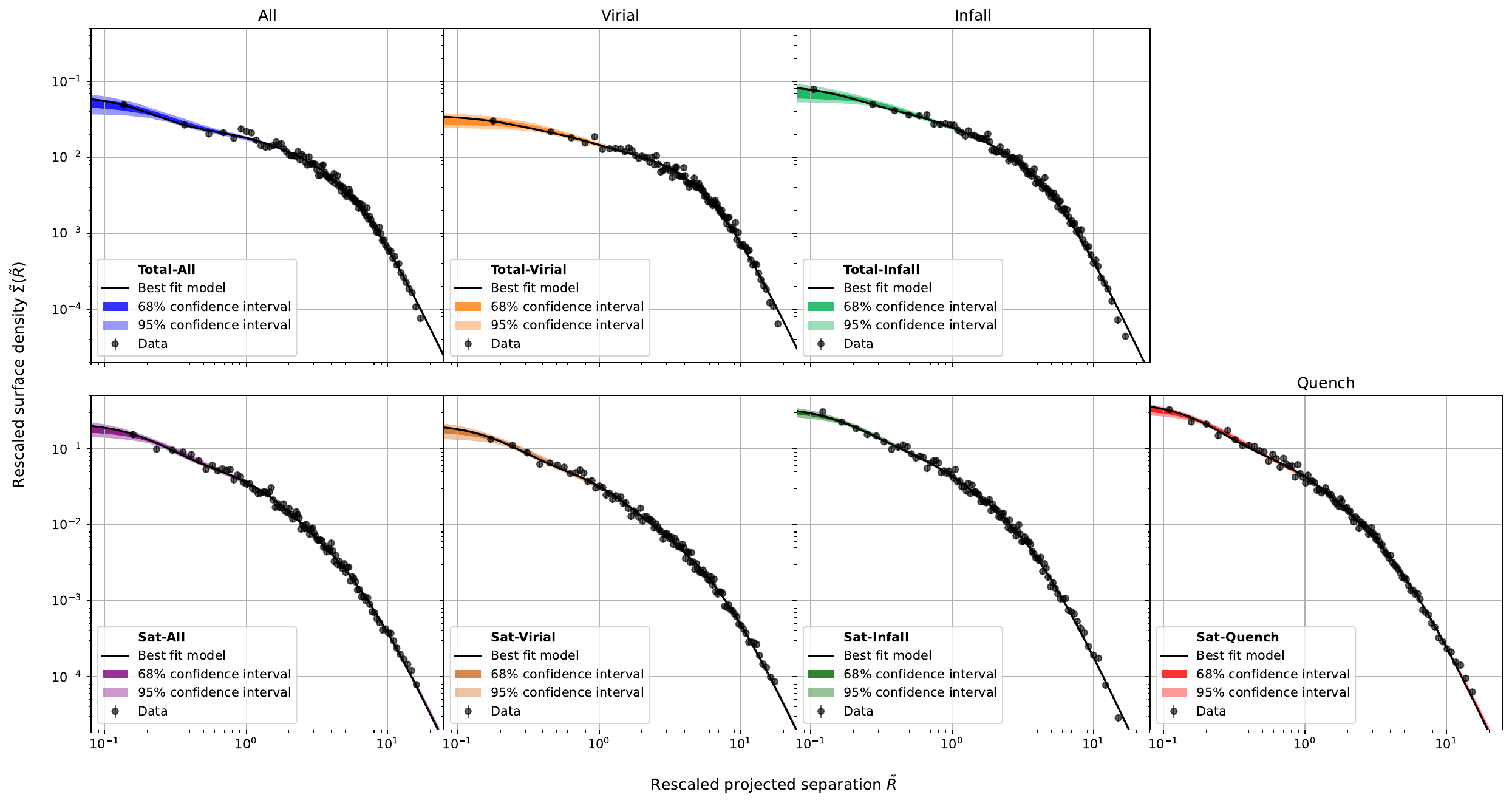}
    \caption{Subhalo surface density profiles for our seven samples. The top row shows the \textit{Total-} samples and the bottom row shows the \textit{Sat-} samples. From left to right the columns show the \textit{-All}, \textit{-Virial}, \textit{-Infall}, and \textit{-Quench} samples. Sample names are also provided in the legend of each panel. Circular black points  represent the measured surface density values (Poisson uncertainties are too small to be seen). The solid black lines represent the best fitting Plummer sphere models (\autoref{eq:Plummer SD}), whilst the coloured shaded regions are the corresponding $68\%$ and $95\%$ confidence intervals. The Plummer sphere model appears to be an appropriate description for the fitted data in each sample.}
    \label{fig:surfden}
\end{figure*}

The surface density fits are shown in \autoref{fig:surfden}. Subhalos in the \textit{Total-} samples show flatter inner surface density profiles and tend to be found at larger $\tilde{R}$ than their \textit{Sat-} counterparts.

\subsubsection{Velocity dispersions and fourth order moments}\label{subsubsec:Methods vel disp calc}

\begin{figure*}
	\includegraphics[width=\textwidth]{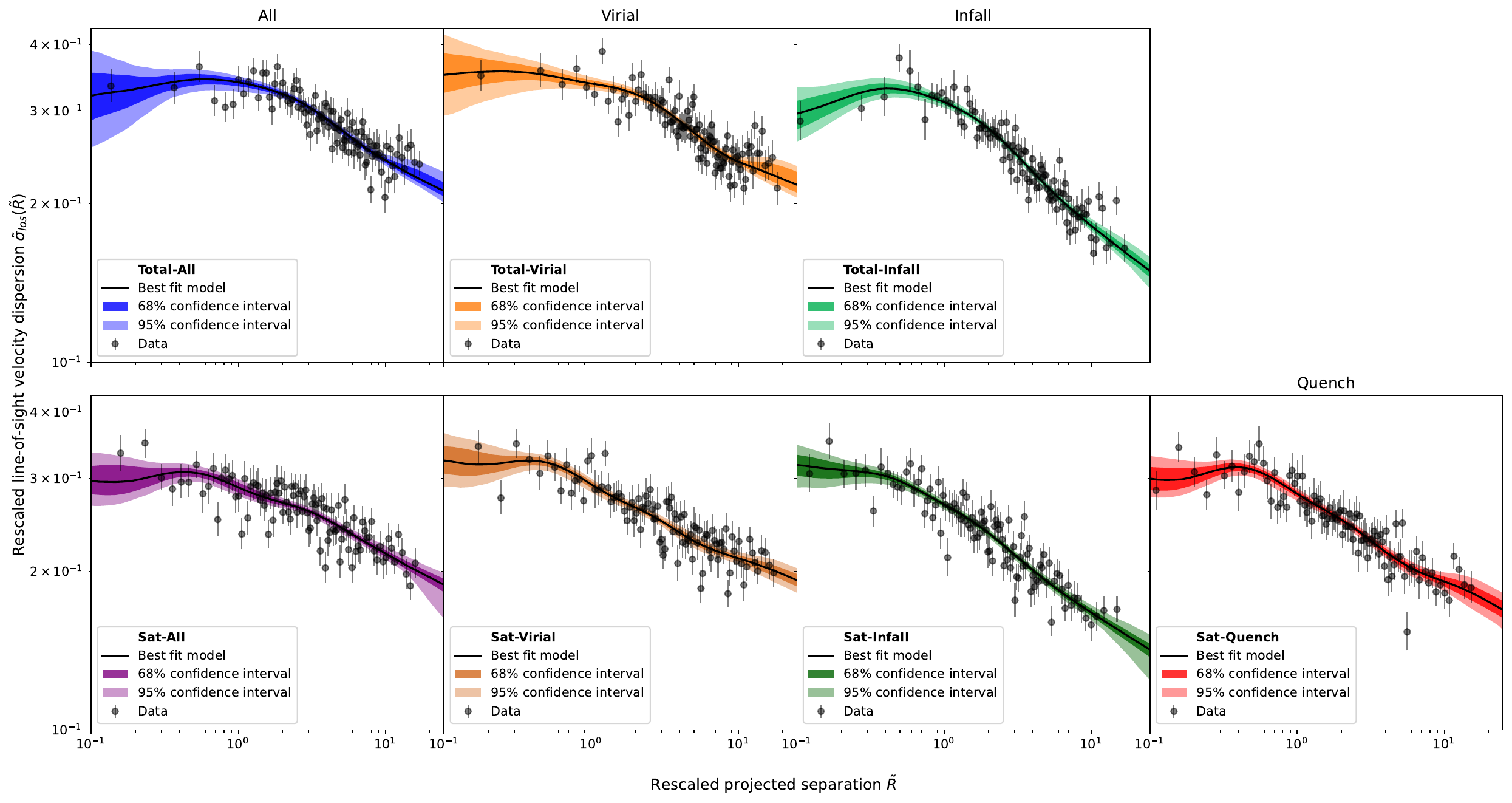}
    \caption{Subhalo line-of-sight velocity dispersion profiles. The top row shows the \textit{Total-} samples and the bottom row shows the \textit{Sat-} samples. From left to right the columns show the \textit{-All}, \textit{-Virial}, \textit{-Infall}, and \textit{-Quench} samples. Sample names are also provided in the legend of each panel. Circular black points with errorbars represent the measured velocity dispersions and their uncertainties. The solid black lines represent the best fit dynamical mass models (defined in \autoref{subsec:Gravsphere}) and the coloured shaded regions are the corresponding $68\%$ and $95\%$ confidence intervals.}
    \label{fig:Gravsphere LOS velocity dispersion}
\end{figure*}

\textsc{Binulator} measures velocity dispersions and fourth order moments by fitting a modified Gaussian to the velocity distribution of each kinematic bin. The motivation provided by \citet{Gravsphere_binulator} is that this allows the velocity distribution to be convolved with the velocity uncertainties (uncertainties can bias the measured second and fourth moments high) and is also more robust for small numbers of tracers. This modified Gaussian has the form
\begin{equation}\label{eq:Modified gauss}
    p_{i}=\frac{\gamma_{v}}{2\alpha_{v}\Gamma\left(\frac{1}{\gamma_{v}}\right)}\exp\left[-\left(\frac{|v_{los,i} -\mu|}{\alpha_{v}}\right)^{\gamma_{v}} \right],
\end{equation}
with $p_{i}$ representing the probability of measuring a line-of-sight velocity of $v_{los,i}$ for the $i$th tracer in the bin. $\Gamma$ is the Gamma function and $\mu$ represents the mean of the distribution. $\alpha_{v}$ characterises the width of the distribution and is related to velocity dispersion by
\begin{equation}
\sigma_{los}^{2} \equiv \alpha_{v}^{2}\Gamma\left(\frac{3}{\gamma_{v}}\right) /\Gamma\left(\frac{1}{\gamma_{v}} \right),
\end{equation}
whilst $\gamma_{v}$ characterises the 'peakiness' of the distribution and is related to kurtosis by
\begin{equation}
    \kappa \equiv \Gamma\left(\frac{5}{\gamma_{v}} \right)\Gamma\left(\frac{1}{\gamma_{v}} \right) / \Gamma\left(\frac{3}{\gamma_{v}} \right)^{2}.
\end{equation}
A pure Gaussian corresponds to the $\gamma_{v}=2$ case. The fourth-order velocity moment is then defined as
\begin{equation}\label{eq:kurtosis 4th order}
    \langle v_{los}^{4} \rangle \equiv \kappa \sigma_{los}^{4}.
\end{equation}

This modified Gaussian is convolved with the error probability distribution function of each tracer using an analytic approximation. This accounts for the velocity uncertainty of each tracer when performing the fit and is defined such that
\begin{equation}\label{eq:Vel disp convolve}
    \alpha^{2}=\alpha_{v}^{2} + \sigma_{e,i}^{2}\Gamma\left(\frac{1}{\gamma_{v}}\right) / \Gamma\left(\frac{3}{\gamma_{v}}\right),
\end{equation}
where $\sigma_{e,i}$ is the velocity uncertainty.

The modified Gaussian is fitted to the velocity data in each bin with \textsc{emcee} using a likelihood of the form
\begin{equation}
    \mathcal{L}_{v} = \prod_{i}p_{i}.
\end{equation}
We use 500 walkers, 10,000 models per walker and a burn-in of 75\% for each bin (unchanged from the standard installation of \textsc{binulator}). There are six parameters per bin to fit, with priors set as: $-1 < \mu < 1$, $0 < \alpha < 1$, $1 < \gamma_{v} < 5$, $10^{-5}<A_{e}<10^{-4}$, $-1 < \mu_{e} < 1$, and $0 < \sigma_{e} < 1$, where $A_{e}$ and $\mu_{e}$ are the amplitude and mean of the velocity uncertainty PDF respectively. \textsc{binulator} priors can be viewed in \autoref{tab:Binulator}.

\begin{figure*}
\includegraphics[width=\textwidth]{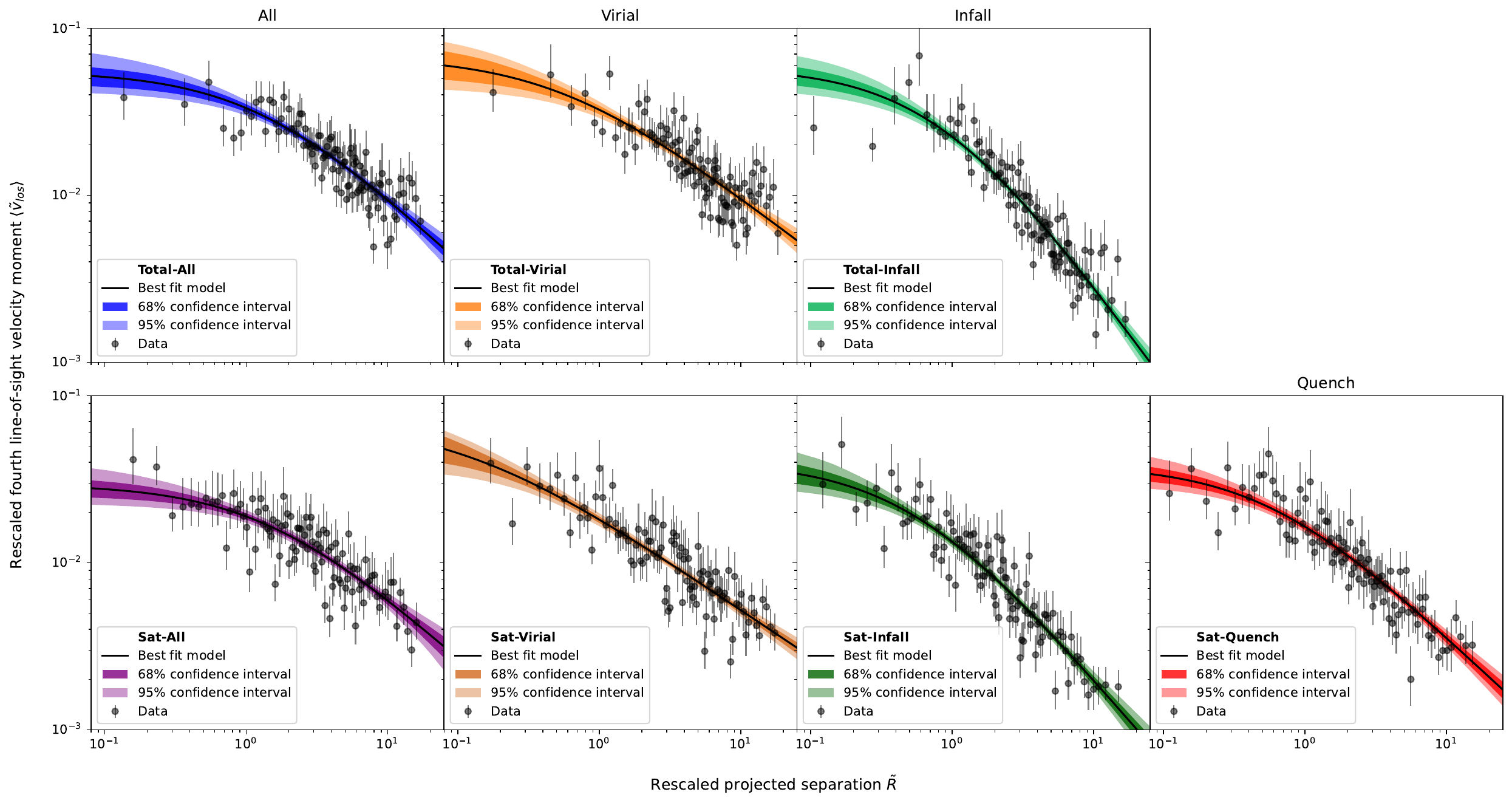}
    \caption{Subhalo $\vzfour$ profiles. The top row shows the \textit{Total-} samples and the bottom row shows the \textit{Sat-} samples. From left to right the columns show the \textit{-All}, \textit{-Virial}, \textit{-Infall}, and \textit{-Quench} samples. Sample names are also provided in the legend of each panel. Circular black points with error bars represent the measured $\langle v_{los}^{4} \rangle$ values and their uncertainties. The solid black lines represent best fitting broken power law models (\autoref{eq:Broken power law}). The coloured shaded regions are the corresponding $68\%$ and $95\%$ confidence intervals. The broken power law model seems to be a good description for the fitted data in each sample.}
    \label{fig:vzfour}
\end{figure*}

We show the $\tilde{\sigma}_{los}^{2}$ and $\vzfour$ profiles in \autoref{fig:Gravsphere LOS velocity dispersion} and \autoref{fig:vzfour} respectively. The uncertainties on $\tilde{\sigma}_{los}^{2}$ are assumed to be Gaussian, whilst those for $\vzfour$ are measured by directly sampling the PDFs \citep[as in][]{Gravsphere_binulator}. The best fitting models shown are described later in \autoref{subsec:Gravsphere} for $\tilde{\sigma}_{los}^{2}$ and \autoref{subsubsec:virial shape binulator} for $\vzfour$).

Comparing samples, the velocity dispersions in \autoref{fig:Gravsphere LOS velocity dispersion} are generally larger for the \textit{Total-} samples when compared to their \textit{Sat-} counterparts. The \textit{-All} and \textit{-Virial} samples also possess larger velocity dispersions than the \textit{-Infall} and \textit{-Quench} samples. This trend is also present for the fourth velocity moments in \autoref{fig:vzfour}.

\subsubsection{Virial shape parameters}\label{subsubsec:virial shape binulator}

Once the surface density and velocity distributions have been fitted, the virial shape parameters can be calculated using
\begin{equation}\label{eq:vs1 los only}
    \tilde{v}_{s1} = \int_{0}^{\infty} \tilde{\Sigma}(\tilde{R}) \vzfour \tilde{R} \mathrm{d}\tilde{R},
\end{equation}
\begin{equation}\label{eq:vs2 los only}
    \tilde{v}_{s2} = \int_{0}^{\infty} \tilde{\Sigma}(\tilde{R}) \vzfour \tilde{R}^{3} \mathrm{d}\tilde{R}.
\end{equation}

However, our data (and any realistic dataset), cannot probe directly to the upper bound of $R=+\infty$. Simply integrating up to the maximum available $R$ is a poor solution to this issue, as we find that the integrands in \autoref{eq:vs1 los only} and \autoref{eq:vs2 los only} do not approach zero by this point and so the integrals will not be converging. For the surface densities, we therefore extrapolate using the Plummer sphere model ($\tilde{\Sigma}_{p}(\tilde{R})$) defined by \autoref{eq:Plummer SD}. For the fourth-order velocity moments, we fit a continuous model, similar to \citet{Gravpshere_2}, and extrapolate to large $\tilde{R}$. We use a broken power law model of the form
\begin{equation}\label{eq:Broken power law}
    \vzfour(\tilde{R}) = A \left( 1 + \frac{\tilde{R}}{R_{4}} \right)^{-k}.
\end{equation}
This model is flat at small radii with $A$ representing the y-intercept. At some scale radius $R_{4}$, the model then transitions to a power law with gradient $-k$. The best fitting models to the $\vzfour$ profiles and their confidence intervals are shown in \autoref{fig:vzfour}. The error bars on the data points are calculated directly from the marginalised $\vzfour$ PDFs. As can be seen, this broken power law model provides a good fit to the available data.

\textsc{binulator} calculates the virial shape PDFs by evaluating \autoref{eq:vs1 los only} and \autoref{eq:vs2 los only} for 2500 random draws of the fitted $\vzfour$ profiles (assuming that the uncertainty on $\tilde{\Sigma}(\tilde{r})$ is negligibly small). We fit continuous distributions to these PDFs to alleviate issues when evaluating likelihoods later on in \autoref{subsubsec: fitting the mass profile}. We find a skew-normal distribution provides a good fit. This is defined as
\begin{equation}\label{eq:Skew-normal}
    SN(x;\mu,\omega,a) = 2\phi(x;\mu,\omega)\Phi(ax;\mu,\omega),
\end{equation}
where $\phi(x;\mu,\omega)$ is a Gaussian PDF evaluated at $x$ with mean $\mu$ and width $\omega$. $\Phi(ax;\mu,\omega)$ is then the corresponding Gaussian cumulative distribution function evaluated at $ax$, with a skewness of $a=0$ for the Gaussian case. These fitted distributions will allow the MCMC chain to freely explore the parameter space without being trapped in (spurious) zero likelihood zones when evaluating mass profiles later in \autoref{subsubsec: fitting the mass profile}.

\subsection{Gravsphere}\label{subsec:Gravsphere}

We measure the $\beta(\tilde{r})$ and $\tilde{M}(<\tilde{r})$ profiles using the \textsc{gravsphere} code \citep{Gravsphere_1, Gravpshere_2, Gravsphere_3}. \textsc{gravsphere} essentially converts the line-of-sight information provided by \textsc{binulator} into 3-D information of the system, mostly through \autoref{eq:Jeans 1-D}. We provide \textsc{gravsphere} with the binned $\tilde{\sigma}_{los}^{2}(R)$, the best fitting surface density parameters $N_{1}, N_{2}, N_{3}, a_{1}, a_{2}, a_{3}$, and the fitted skew-normal distributions of $\tilde{v}_{s1}$ and $\tilde{v}_{s2}$. Below we detail how \textsc{gravsphere} functions, changes we have made to the standard installation, along with the 3-D models we fit the line-of-sight data to.

\subsubsection{Number density}\label{subsec:Number density}

We leave the number density model used by \textsc{gravsphere} unchanged. As mentioned in \autoref{subsubsec:Surface density}, the Plummer sphere surface density model (\autoref{eq:Plummer SD}) has a corresponding analytic number density profile, which takes the form
\begin{equation}\label{eq:Plummer nu}
    \nu(r) = \sum_{s} \frac{3N_{s}}{4\pi a_{s}^{3}} \left(1 + \frac{r^{2}}{a_{s}^{2}} \right)^{-5/2}.
\end{equation}
The $N_{s}$ and $a_{s}$ parameters are the same as those fitted to the surface densities in \autoref{subsubsec:Surface density}.

\citet{Gravsphere_1} explain in their Section 2.2 that these surface density parameters are only allowed to vary within a fixed tolerance of their best fitting values in \textsc{gravsphere}. They justify this as being more efficient, since $\Sigma(R)$ is expected to be well constrained and because it will avoid sampling trivial degeneracies between the individual Plummer spheres. We choose a tolerance of $25\%$ above and below the best fit values as priors, since this is large enough to let the parameters vary across most of their posteriors. \textsc{gravsphere} priors can be viewed in \autoref{tab:Gravsphere}.

\subsubsection{Velocity anisotropy}\label{subsubsec:beta model}

We leave the velocity anisotropy model used by \textsc{gravsphere} unchanged. This has the form
\begin{equation}\label{eq:beta model}
    \beta(r) = \beta_{0} + \frac{\left(\beta_{\infty}  - \beta_{0}\right)}{1 + \left(\frac{r_{0}}{r} \right)^{n}},
\end{equation}
which transitions from $\beta_{0}\equiv\beta(r=0)$ to $\beta_{\infty}\equiv\beta(r=\infty)$ at a radius $r_{0}$ and sharpness $n$. When integrated, this model has an analytic $g(r)$ (needed to evaluate \autoref{eq:Jeans 3D integrated}) of the form
\begin{equation}
    g(r) = r^{2\beta_{\infty}}\left[\left(\frac{r_{0}}{r} \right)^{n} + 1\right]^{\frac{2}{n}\left(\beta_{\infty} - \beta_{0} \right)}.
\end{equation}

Referring back to \autoref{eq:Beta definition}, we see that $\beta_{0}, \beta_{\infty} \in \left[-\infty,1 \right]$. To avoid this infinite parameter space, \citet{Gravsphere_1} define a 'symmetrised anisotropy parameter',
\begin{equation}
    \overline{\beta} \equiv \frac{\beta}{2-\beta}
\end{equation}
which has limits $\overline{\beta}=-1$ for a fully tangential velocity distribution and $\overline{\beta}=+1$ for a fully radial velocity distribution. We set priors on the symmetrised parameters of $-1<\overline{\beta}_{0}<1$, $-1<\overline{\beta}_{\infty}<1$, $0<\log_{10}( r_{0})<\log_{10}(15)$, and $0<n<5$. \textsc{gravsphere} priors can be viewed in \autoref{tab:Gravsphere}.

\subsubsection{Mass model}\label{subsubsec:Mass model}

We also leave the mass profile used by \textsc{gravsphere} unchanged. This is called the cNFWt profile  \citep{Gravpshere_2} and is a modification of the NFW profile that allows for coring in the inner regions (c) and tidal effects at the outer regions (t). We use this model as it is sensitive to any deviations from NFW by design. Although our rescaling procedure allows us to test consistency with the NFW profile, it is not optimised to quantify deviations from this profile on a halo-by-halo basis. In other words, the measured values of the coring and tidal parameters in a stack (where these deviate from NFW) may not necessarily reflect the values of those parameters that describe individual halos.  

The coring is modelled by
\begin{equation}
    M_{cNFW}(<r) = M_{NFW}(<r) f^{n_{c}},
\end{equation}
with 
\begin{equation}
    f^{n_{c}} = \left[\tanh\left(\frac{r}{r_{c}} \right) \right]^{n_{c}}.
\end{equation}
$r_{c}$ controls the radius of the core and $n_{c}$ controls the strength of the coring. For an NFW halo we have  $n_{c}=0$ and $r_{c}$ will be unconstrained. Meanwhile, $n_{c}=1$ would represent a cored halo with a flat inner density profile.

The tidal effects are only modelled at radii larger than the tidal radius $r_{t}$, beyond which the outer slope is modified. For $r>r_{t}$ the mass profile is
\begin{equation}
\begin{aligned}
    M_{cNFWt}(<r) = & M_{cNFW}(<r_{t}) \\
     & + 4\pi\rho_{cNFW}(r_{t}) \frac{r_{t}^{3}}{3-\delta} \left[\left(\frac{r}{r_{t}} \right)^{3-\delta} - 1\right],
\end{aligned}
\end{equation}
where $-\delta$ is the outer logarithmic density gradient. This framework is flexible enough to discern between an infinitely flat rotation curve ($\delta=2$) and an NFW halo ($\delta=3$). 

The final two mass parameters are the NFW virial mass, $M_{vir}$, and concentration, $c_{vir}$. These parameters are dependent on the critical density of the Universe, as they are usually defined by the enclosed mass at which the overdensity is 200 times the critical density. Whilst this definition is useful for identifying individual gravitationally bound systems, it has little physical meaning in our rescaled space (the \textit{rescaled} critical density will vary from halo to halo according to their individual $M_{s}$ and $R_{s}$). For the fitting of rescaled mass profiles, it is therefore more useful to start by adopting an expected value for the virial mass and to then derive the corresponding product of the overdensity and critical density, $\tilde{\Delta \rho_{c}}$. An obvious choice for the rescaled virial mass to begin with is $\tilde{M}_{vir} = 1$. The corresponding fitted concentration for an NFW system can be found by solving \autoref{eq:Mass rescaled} at the rescaled virial radius $\tilde{r}_{vir}$, which by definition will be equal to this fitted concentration and so can be expressed as
\begin{equation}
    \tilde{M}_{vir} = \ln(1 + \tilde{c}_{vir}) - \frac{\tilde{c}_{vir}}{1 + \tilde{c}_{vir}}.
\end{equation}
Solving numerically when $\tilde{M}_{vir} = 1$ yields $\tilde{c}_{vir} \approx 5.3$. The corresponding value of $\tilde{\Delta \rho_{c}}$ can be found by rearranging \autoref{eq:Virial mass X} such that
\begin{equation}
    \tilde{\Delta \rho_{c}} = \frac{3 \tilde{M}_{vir}}{4 \pi \tilde{c}_{vir}^{3}}.
\end{equation}
In practical terms, we therefore set the critical density to unity and the overdensity equal to $\tilde{\rho}_{c} \approx 3/\left(4 \pi \times 5.3^{3}\right)$ within the constants of \textsc{Gravsphere}. Again, the critical density and overdensity have little physical meaning in the rescaled space and the values chosen here are used for fitting purposes within \textsc{Gravsphere} only. This choice provides a clear benchmark to compare the dynamical mass profiles to; we can expect to recover a fitted virial mass of $\tilde{M}_{vir} = 1$ and a fitted concentration of $\tilde{c}_{vir} \approx 5.3$ for NFW systems. 

In total, there are six parameters to fit in the cNFWt model: the two NFW parameters, the two coring parameters, and the two tidal parameters. We set priors of $-1<\log_{10} (M_{vir})<1$, $0.5<c_{vir}<50$, $-1<\log_{10} (r_{c})<\log_{10}(15)$, $-1<n_{c}<1$, $-1<\log_{10} (r_{t})<\log_{10}(15)$, and $0.5 < \delta <4.5$. \textsc{gravsphere} priors can be viewed in \autoref{tab:Gravsphere}.

\subsubsection{Fitting the mass profile}\label{subsubsec: fitting the mass profile}

In total, 16 parameters are allowed to vary in \textsc{gravsphere} when measuring the mass profile: four for the $\beta(\tilde{r})$ profile, six for the $\tilde{M}(<\tilde{r})$ profile, along with the six $\tilde{\Sigma}(\tilde{R})$ parameters (albeit in a more limited way). The likelihood is split into a velocity dispersion component and a virial shape component. 

For the velocity dispersions, the model parameters are used to calculate the model velocity dispersion $\sigma_{model}^{2}(R)$ through \autoref{eq:Jeans 1-D}. The log-likelihood is then calculated as
\begin{equation}\label{eq:likelihood vel disp}
    \ln(\mathcal{L}_{\sigma}) = -\frac{1}{2}\sum_{i}\frac{\left[ \sigma_{los}^{2}(R_{i}) - \sigma_{model}^{2}(R_{i})\right]^{2}}{\alpha_{\sigma^{2},i}^{2}},
\end{equation}
where $R_{i}$ is the projected separation of the $i$th kinematic bin and $\alpha_{\sigma^{2},i}$ is the Gaussian uncertainty on the line-of-sight velocity dispersion. To accommodate our rescalings, we set $G=1$ within the constants of \textsc{Gravsphere} for these calculations (this is the correct value in the rescaled space by definition).

For the virial shape parameters, the model values $v_{s1, model}$ and $v_{s2,model}$ are calculated by feeding the model parameters into the $r$-integrals of \autoref{eq:4th order beta 1} and \autoref{eq:4th order beta 2} respectively. We modified this likelihood function to use a fitted skew-normal distribution rather than sampling the virial shape PDFs directly (see \autoref{subsubsec:virial shape binulator}). The likelihood is calculated by evaluating these fitted distributions (normalised such that the maximum height is 1) at the model values. The log-likelihood is therefore
\begin{equation}
    \ln(\mathcal{L}_{vsp}) = \sum_{i=1}^{2} \ln\left[ \frac{SN(v_{si,model};\mu_{i},\omega_{i},a_{i})}{\max \left[ SN(x;\mu_{i},\omega_{i},a_{i})\right]} \right],
\end{equation}
with the skew-normal distribution defined in \autoref{eq:Skew-normal}.

The total log-likelihood supplied to \textsc{emcee} is the sum of these two components,
\begin{equation}
    \ln(\mathcal{L}_{tot}) = \ln(\mathcal{L}_{\sigma}) + \ln(\mathcal{L}_{vsp}).
\end{equation}
We use 250 walkers, 60,000 models per walker and a burn-in of $25\%$ to sample the posteriors and find the best fitting parameters.

\subsection{Summary of parameters and priors}\label{subsec:parameters summary}

We summarise our parameters and priors in \autoref{tab:Binulator} for \textsc{binulator} and \autoref{tab:Gravsphere} for \textsc{gravsphere}. The same priors are used for every sample. We note here that \textsc{gravsphere} includes functionality to model additional mass/kinematic components. Since we are only interested in the total mass profile, these are disabled in our analysis.

\begin{table}
    \centering
    \caption{Summary of model parameters and prior bounds used in \textsc{binulator}. We use the same priors for each sample.}
    \label{tab:Binulator}
    \begin{tabular}{lcc}
        \hline
        \multicolumn{3}{c}{Binulator} \\
        Parameter & Lower Bound & Upper Bound \\
        \hline
        \multicolumn{3}{l}{\textit{Surface density parameters}} \\
        $N_{1}$ & $10^{-4}$ & $10^{2}$ \\
        $N_{2}$ & $10^{-4}$ & $10^{2}$ \\
        $N_{3}$ & $10^{-4}$ & $10^{2}$ \\
        $a_{1}$ & $0.25$ & $3$ \\
        $a_{2}$ & $0.25$ & $3$ \\
        $a_{3}$ & $0.25$ & $10$ \\
        \\[0.5em]
        \multicolumn{3}{l}{\textit{Velocity bin parameters}} \\
        $\mu_{v}$ & $-1$ & $1$ \\
        $\alpha_{v}$ & $0$ & $1$ \\
        $\gamma_{v}$ & $1$ & $5$ \\
        $A_{e}$ & $10^{-5}$ & $10^{-4}$ \\
        $\mu_{e}$ & $-1$ & $1$ \\
        $\sigma_{e}$ & $0$ & $1$ \\
        \hline
    \end{tabular}
\end{table}

\begin{table}
    \centering
    \caption{Summary of model parameters and prior bounds used in \textsc{gravsphere}. The subscript $best$ represents the best fitting values measured by \textsc{binulator}. We use the same priors for each sample.}
    \label{tab:Gravsphere}
    \begin{tabular}{lcc}
        \hline
        \multicolumn{3}{c}{Gravsphere} \\
        Parameter & Lower Bound & Upper Bound \\
        \hline
        \multicolumn{3}{l}{\textit{Number density parameters}} \\
        $N_{1}$ & $0.75 N_{1,best}$ & $1.25 N_{1,best}$ \\
        $N_{2}$ & $0.75 N_{2,best}$ & $1.25 N_{2,best}$ \\
        $N_{3}$ & $0.75 N_{3,best}$ & $1.25 N_{3,best}$ \\
        $a_{1}$ & $0.75 a_{1,best}$ & $1.25 a_{1,best}$ \\
        $a_{2}$ & $0.75 a_{2,best}$ & $1.25 a_{2,best}$ \\
        $a_{3}$ & $0.75 a_{3,best}$ & $1.25 a_{3,best}$ \\
        \\[0.5em]
        \multicolumn{3}{l}{\textit{Symmetrised velocity anisotropy parameters}} \\
        $\overline{\beta}_{0}$ & $-1$ & $1$ \\
        $\overline{\beta}_{\infty}$ & $-1$ & $1$ \\
        $\log_{10} (r_{0})$ & $0$ & $\log_{10} (15)$ \\
        $n$ & $0$ & $5$ \\
        \\[0.5em]
        \multicolumn{3}{l}{\textit{Mass profile parameters}} \\
        $\log_{10} (M_{vir})$ & $-1$ & $1$ \\
        $c_{vir}$ & $0.5$ & $50$ \\
        $\log_{10} (r_{c})$ & $-1$ & $\log_{10}(15)$ \\
        $n_{c}$ & $-1$ & $1$ \\
        $\log_{10} (r_{t})$ & $-1$ & $\log_{10} (15)$ \\
        $\delta$ & $0.5$ & $4.5$ \\
        \hline
    \end{tabular}
\end{table}

\section{Results}\label{sec:results} 

We show the dynamical mass profiles and velocity anisotropy profiles recovered from \textsc{gravsphere} below, in \autoref{subsec:Results Mass profile} and \autoref{subsec:Results beta} respectively.

\subsection{Dynamical mass profile}\label{subsec:Results Mass profile}

\begin{figure*}
\centering
	\includegraphics[width=\textwidth]{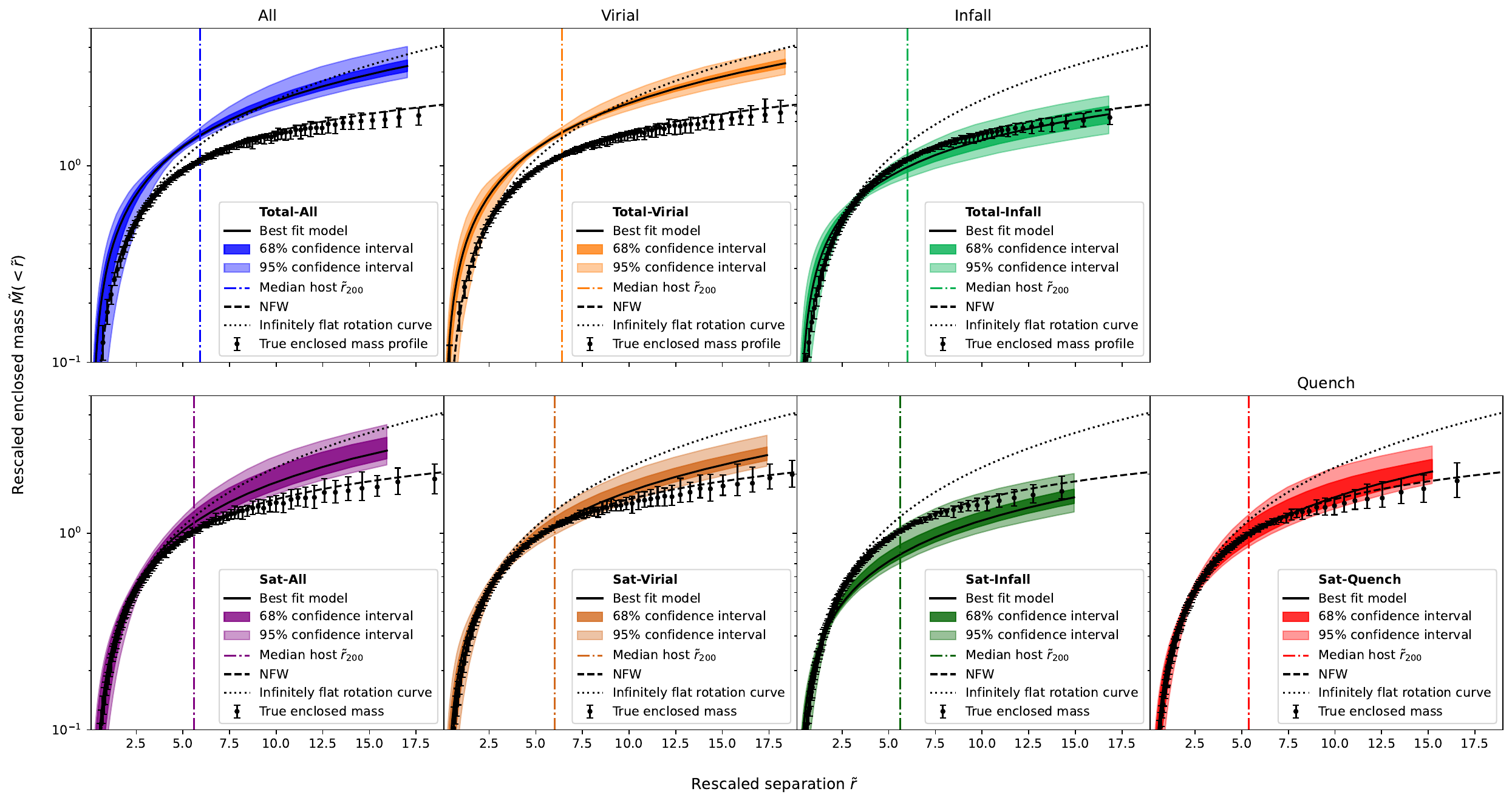}
    \caption{Dynamical mass profiles recovered from the line-of-sight data. The top row shows the \textit{Total-} samples and the bottom row shows the \textit{Sat-} samples. From left to right the columns show the \textit{-All}, \textit{-Virial}, \textit{-Infall}, and \textit{-Quench} samples. Sample names are also provided in the legend of each panel. Circular black points with errorbars represent the median true mass profile and the $1\sigma$ scatter about these medians. The solid black lines represent the best fitting dynamical mass profile models (\autoref{subsubsec:Mass model}). The coloured shaded regions are the corresponding $68\%$ and $95\%$ confidence intervals. Also included in each panel as a reference are black dashed lines to show the rescaled NFW profile (\autoref{eq:Mass rescaled}), black dotted lines to show the infinitely flat rotation curve model defined in \autoref{eq:Mass flat RC}, and vertical coloured dash-dotted lines to show the median host $\tilde{r}_{200}$. On the top row, the recovered mass profiles are biased high for the \textit{Total-All} and \textit{Total-Virial} samples, even appearing consistent with the flat rotation curve model. The \textit{Total-Infall} sample recovers an unbiased mass profile at all radii (up to $\sim2.5$ times the median $\tilde{r}_{200}$). On the bottom row, the recovered mass profiles are biased high for the \textit{Sat-All} and \textit{Sat-Virial} samples beyond the median $\tilde{r}_{200}$ and this bias is smaller than their \textit{Total-} counterparts. The recovered mass profile is biased low for the \textit{Sat-Infall} sample beyond $\sim0.25$ times the median $\tilde{r}_{200}$. The \textit{Sat-Quench} sample recovers an accurate mass profile up to $\sim1.5$ times the median $\tilde{r}_{200}$ and slightly overestimates the mass beyond this point. This overestimate is smaller in magnitude than for the \textit{Sat-All} and \textit{Sat-Virial} samples.}
    \label{fig:Mass profile Gravsphere}
\end{figure*}

We display the recovered dynamical mass profiles in \autoref{fig:Mass profile Gravsphere}. Median true mass profiles are recalculated for each sample, in the same manner as for \autoref{fig:Rescale true profile} using only the subhalos within that sample. We include as a reference in each panel the median host virial radius of the sample, the predicted NFW mass profile (\autoref{eq:Mass rescaled}) and the infinitely flat rotation curve model defined in \autoref{eq:Mass flat RC}.

First, looking at the \textit{Total-} samples (which will contain many dark matter only objects), we see that the \textit{Total-All} dynamical mass profile is biased high and even appears consistent with the infinitely flat rotation curve model. Selecting for relaxed groups with the \textit{Total-Virial} sample provides little improvement to this bias. In both cases, the true mass profile is not recovered. In contrast, selecting for subhalos which infell long ago with \textit{Total-Infall} produces a dynamical mass profile that recovers the true profile without bias at large radii, up to 2.5 times the median host $\tilde{r}_{200}$.

Looking now at the \textit{Sat-} samples (which should correlate closer to satellite galaxies that could actually be observed), we similarly see that the \textit{Sat-All} sample dynamical mass profile is biased high, although the bias in this case is not as significant as for the \textit{Total-All} sample. Selecting for relaxed groups with the \textit{Sat-Virial} sample provides some improvement in this case. The \textit{Sat-Virial} sample is consistent with the true profile to just beyond the median host $\tilde{r}_{200}$, before becoming biased high. Again though, the bias is smaller than for its \textit{Total-Virial} counterpart. Selecting for satellites which infell long ago with \textit{Sat-Infall} surprisingly leads to an underestimate of the true mass profile, though the discrepancy is $\lesssim 2 \sigma$ at all separations probed. Selecting for quenched satellites with \textit{Sat-Quench} provides the best recovery in this case, with an accurate recovery of the mass profile up to $\sim1.5$ times the median $\tilde{r}_{200}$ before becoming biased high at larger radii. This bias is smaller in magnitude than for the \textit{Sat-All} and \textit{Sat-Virial} samples. However, the uncertainties are also too large to rule out the infinitely flat rotation curve model definitively (i.e. at $>5 \sigma$ significance). It is also noteworthy that \textit{Sat-Infall} and \textit{Sat-Quench} provide different results, considering that quenched satellites should also be those that infell long ago. This discrepancy may be caused by the relative bias to high host group mass for the quenched satellites (see \autoref{fig:Group mass hist}). However, other sources of bias are clearly present in the other samples as well and we discuss these further in \autoref{subsec:correction terms}.

It is somewhat surprising that the analysis on subhalos is so misleading; based purely off the \textit{Total-} samples we would expect quality cuts based on infall time to allow the most accurate recovery of the mass profile and that selecting for relaxed groups would produce results incompatible with the truth. Instead, when applied to satellite galaxies in the \textit{Sat-} samples, the infall time cut produces an underestimate of the mass and the selection of relaxed groups may even be preferable. Still, it is reassuring that selecting for quenched satellites produces the best recovery in this case, as this is the simplest quality cut to apply to real data (apart from applying no cuts at all like the \textit{-All} biased samples).

\subsection{Velocity anisotropy}\label{subsec:Results beta}

\begin{figure*}
	\includegraphics[width=\textwidth]{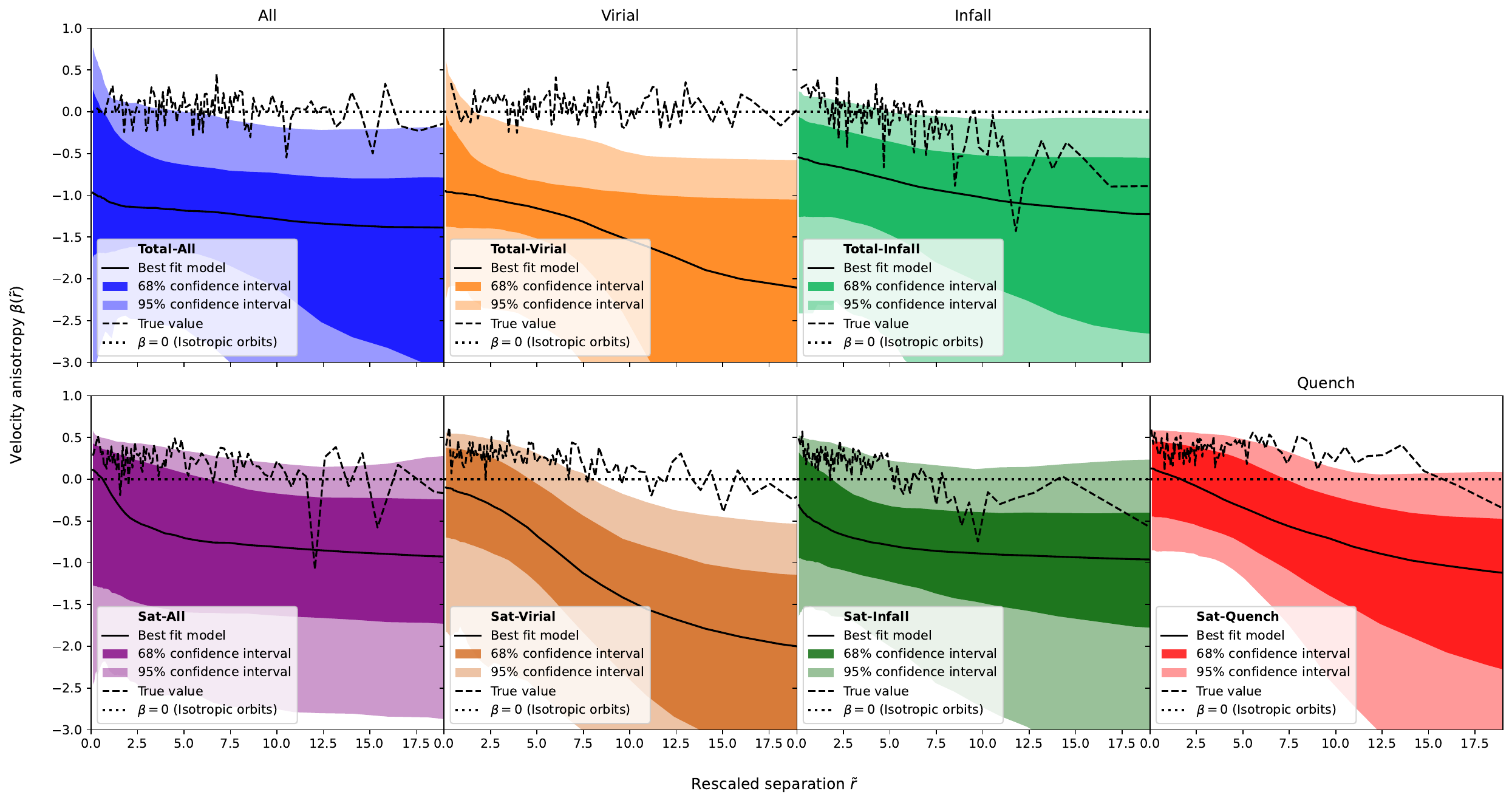}
    \caption{Velocity anisotropy profiles recovered from the line-of-sight data. The top row shows the \textit{Total-} samples and the bottom row shows the \textit{Sat-} samples. From left to right the columns show the \textit{-All}, \textit{-Virial}, \textit{-Infall}, and \textit{-Quench} samples. Sample names are also provided in the legend of each panel. The solid black lines represent the best fitting velocity anisotropy models (\autoref{subsubsec:beta model}). The coloured shaded regions are the corresponding $68\%$ and $95\%$ confidence intervals. Black dashed lines show the true velocity anisotropy profiles, calculated directly using the 3-D data. As a reference, a black dotted line is shown at $\beta =0$ in each panel, which represents the case of isotropic orbits. For all 7 samples the recovered velocity anisotropy is biased towards negative values. The severity of this bias differs between samples. The best fitting models \textit{Total-} models are generally beyond $\gtrsim 1\sigma$ at all radii, whilst the best fitting \textit{Sat-} models are within $\lesssim 1 \sigma$ of the true value at low radii, but exceed this threshold beyond $\tilde{r} \gtrsim 5$.}
    \label{fig:beta normal}
\end{figure*}

We show the $\beta(\tilde{r})$ profiles recovered from line-of-sight data in \autoref{fig:beta normal}. The true values calculated directly from the 3-D data are also shown.

For each sample the recovered velocity anisotropies are biased low. For the \textit{Sat-All}, \textit{Sat-Virial}, and \textit{Sat-Quench} samples the recovered profiles are within $\lesssim 1 \sigma$ at the innermost radii only and exceed this threshold beyond $\tilde{r}\gtrsim5$. Consulting \autoref{eq:4th order beta 1} and \autoref{eq:4th order beta 2}, an underestimate in velocity anisotropy indicates an overestimate in the virial shape parameters. In turn, this implies that the fourth velocity moments may have been overestimated. This could be an indication that the interloper removal in \autoref{subsec:Interlopers} was insufficient. Alternatively, it could also be caused by systematic effects. Without further improvement, these low recovered velocity anisotropies would suggest that satellites are on tangentially biased orbits, when the true values indicate they are actually on more isotropic orbits.

These biased velocity anisotropies indicate that it is important to further explore potential systematic effects before any application to real data as the dynamical mass profiles may also be subject to change. From \autoref{eq:Jeans 3D} and \autoref{eq:Jeans 1-D}, we may naively expect that a negative bias in $\beta$ would cause a positive bias in $M(<r)$, although the source of the bias itself may have further effects. We discuss potential systematics in \autoref{subsec:correction terms}.

\section{Discussion}\label{sec:discussion}

\subsection{Potential systematic effects}\label{subsec:correction terms}

Our results indicate that dynamical modelling of rescaled and stacked data should allow for an accurate recovery of galaxy group mass profiles to large radii when selecting for quenched satellites. However, biases in the recovered velocity anisotropies are worrying and may indicate the presence of unaccounted for systematic effects. Being able to quantify these effects would improve confidence in this methodology and may allow us to extend accurate mass profile recovery to even larger radii. This in turn would provide a stronger test of the NFW profile in galaxy groups.

We highlight here five potential sources of bias at large radii: the effect of the background cosmology, acceleration of the reference frame, bulk radial velocities, bulk tangential velocities and the two-halo term. The magnitude of these effects has been explored for galaxy clusters, but not yet for galaxy groups. We will discuss most of these effects qualitatively here, as we plan to fully quantify them in a follow-up paper. The bulk tangential motions are an exception and we will provide an argument for why they can be ruled out for stacked systems in \autoref{subsubsec:tangential motion}.

\subsubsection{Cosmological effects}
Firstly, cosmological effects have been explored by \citet{Falco_JeansSwindle_2013,Falco_ModifiedJeans_2013}, who find that the repulsive action of dark energy is dominant over the attractive action of the background matter density. The background cosmology therefore contributes a negative component to the gravitational potential. The background cosmology also induces a radial outflow of tracers from the system, which can be modelled in the Jeans equation by considering the radial and time evolution of the mean radial velocity. The direct contribution of the background cosmology to the gravitational potential is cancelled out by its contribution to the mean radial velocities. Background cosmology therefore ends up inducing no bias in the Jeans equation \footnote{This is often referred to as the `Jeans swindle'}.

\subsubsection{Frame acceleration}
Next, the frame acceleration effect has been explored in galaxy clusters by \citet{Cai_2025}. They describe that the Jeans equation assumes an inertial reference frame, which should correlate to a frame with no peculiar acceleration in a cosmological context. For galaxy groups, we would measure satellite line-of-sight velocities using the central as the reference frame. These centrals will be free-falling with respect to the halo and the halo itself is likely to be in free-fall towards local cosmic structure. The question then becomes whether these free-fall accelerations are significant enough to bias the recovered mass profile. For a typical galaxy group we expect the central to completely dominate the system and within this view the central's halo is essentially the group halo. We may therefore expect the central to have a negligible acceleration with respect to the dark matter halo. It is less clear whether the peculiar acceleration of the group itself will be more significant, although selecting for isolated groups (which are therefore not near any structures to accelerate onto) would provide a solution if this is an issue.

\subsubsection{Bulk radial motion}
For the bulk radial velocities, there will be a contribution that must be accounted for in the 3-D Jeans equation and a second contribution during the projection to the line-of-sight velocity dispersions/fourth velocity moments. Bulk radial motions are modelled for the 3-D Jeans equation by \citet{Falco_ModifiedJeans_2013}, but the projection to the line-of-sight is not explored. Satellites accreting onto a group will induce a bulk radial motion at large radii, so this is a potential source of the biases in our results.

\subsubsection{Bulk tangential motion}\label{subsubsec:tangential motion}
We can rule bulk tangential motion as a source of bias for stacked systems by considering how the velocity moments are defined. Bulk tangential motion is typically expected to induce a bias as the standard Jeans equation (\autoref{eq:Jeans 3D}) assumes a dispersion supported system, so any rotational support is not modelled. This can lead to an underestimate of the support against gravitational collapse within a system. A more general form \citep[e.g. see][]{Falco_ModifiedJeans_2013} can be given by
\begin{equation}\label{eq:General Jeans}
    \frac{1}{\nu}\frac{\partial}{\partial r} \left[\nu \langle v_{r}^{2}\rangle \right] + \frac{2 \langle v_{r}^{2} \rangle - \langle v_{\theta}^{2}\rangle - \langle v_{\phi}^{2} \rangle}{r} = -\frac{GM(<r)}{r^{2}}.
\end{equation}
The velocity dispersion $\sigma$ is related to the second velocity moment $\langle v^{2} \rangle$ and the mean velocity $\overline{v}$ through
\begin{equation}
    \sigma^{2} \equiv \langle v^{2} \rangle - \overline{v}^{2},
\end{equation}
so the standard Jeans equation is recovered when all mean velocities are zero in \autoref{eq:General Jeans}.

Next, we can consider a system where all tracers are on purely tangential orbits, such that $\langle v_{r} \rangle = \sigma_{r} = 0$. In this case, \autoref{eq:General Jeans} can be reduced to
\begin{equation}\label{eq:Jeans tangential}
    \frac{\overline{v}_{\theta}^{2} + \sigma_{\theta}^{2} + \overline{v}_{\phi}^{2} + \sigma_{\phi}^{2}}{r} \equiv \frac{\overline{v}_{t}^{2} + \sigma_{t}^{2}}{r} = \frac{GM(<r)}{r^{2}},
\end{equation}
as $\overline{v}_{t}^{2} = \overline{v}_{\theta}^{2} + \overline{v}_{\phi}^{2}$ and $\sigma_{t}^{2} = \sigma_{\theta}^{2} + \sigma_{\phi}^{2}$. For an angle $\psi$ between a tracer's polar and tangential velocities, we will have
\begin{align}
    v_{t} \cos \psi &= v_{\theta} \\
    v_{t} \sin \psi &= v_{\phi},
\end{align}
with corresponding moments of
\begin{align}
    \langle v_{\theta}^{n} \rangle &= \langle v_{t}^{n} \rangle \langle \cos^{n}\psi \rangle \label{eq:vtheta moments}\\
    \langle v_{\phi}^{n} \rangle &= \langle v_{t}^{n} \rangle \langle \sin^{n}\psi \rangle. \label{eq:vphi moments}
\end{align}
The average of some angular function $f(\psi)$ is dependent on the probability distribution of angles $P(\psi)$ and is given by
\begin{equation}\label{eq:angular average}
    \langle f(\psi)\rangle = \int_{0}^{2\pi} P(\psi) f(\psi) d\psi.
\end{equation}
So, the polar and azimuthal velocity moments are dependent on the distribution of $\psi$.

Now let's consider a spiral galaxy like system, where we will assume the tracers are all characterised by an equal tangential velocity $v_{t} = V_{c}$. If all tracers move in the positive azimuthal direction, then the angles $\psi$ are distributed as the delta function $P(\psi) = \delta(\psi - \frac{\pi}{2})$. Using \autoref{eq:vtheta moments}, \autoref{eq:vphi moments}, and \autoref{eq:angular average} to evaluate the first and second velocity moments we find that $\overline{v}_{\phi} = V_{c}$ and $\overline{v}_{\theta} = \langle v_{\theta}^{2} \rangle = \langle v_{\phi}^{2} \rangle = 0$. \autoref{eq:Jeans tangential} then evaluates to $\frac{V_{c}^{2}}{r} = \frac{GM(<r)}{r^{2}}$ (as would be expected).

Finally, let's consider a stack of many such spiral galaxy like systems, randomly oriented with respect to each other. Whereas $P(\psi)$ is described by a delta function for an individual galaxy, it will instead become a uniform distribution when stacking many randomly oriented galaxies, so becomes $P(\psi) = \frac{1}{2\pi}$. Once again evaluating the first and second velocity moments shows that $\overline{v}_{\theta} = \overline{v}_{\phi} = 0$ and $\langle v_{\theta}^{2} \rangle = \sigma_{\theta}^{2} = \langle v_{\phi}^{2} \rangle = \sigma_{\phi}^{2} = \frac{1}{2}V_{c}^{2}$. \autoref{eq:Jeans tangential} then again evaluates to the correct result of $\frac{V_{c}^{2}}{r} = \frac{GM(<r)}{r^{2}}$, even though the mean tangential velocity of the stacked data is zero!

To summarise then, we show here that bulk tangential motion should not induce any systematic effects when applying the Jeans equation to stacked systems. It is the second tangential velocity moment $\langle v_{t}^{2} \rangle$ only that contributes and this is preserved during the stacking process as the velocity dispersion of the stacked system absorbs the rotational support of the individual systems. No correction needs to be made as the tangential motion is already accounted for. Notably, the same will not be true for radial motion as the mean radial velocity $\overline{v}_{r}$ has contributions independent of the second radial velocity moment \citep[see][]{Falco_ModifiedJeans_2013}.

\subsubsection{Two-halo term}
The final source of bias to discuss is the two-halo term, which accounts for the mass contribution of nearby structure and has already been explored in some of the galaxy-galaxy weak lensing studies \citep{Mistele_2024}. In clusters, this is expected to cause deviations above NFW in the true mass profile at around 2 virial radii \citep{Oguri_2011}. The outermost bins in our analysis approach this range, so this is another possible source of the bias within our results.

\subsection{Considerations before observational application}

The work we present here represents an idealised case where various group and subhalo properties are known perfectly from the simulation data. It is therefore useful to consider qualitatively whether proxies for such information are theoretically accessible within observational data and what challenges further observational uncertainties may introduce.

\subsubsection{Concentrations}\label{subsubsec:discussion concentrations}

One immediate area of concern is whether it is realistic to expect accurate concentrations to be available for real galaxy groups. As discussed in \autoref{subsec:Rescaling}, some estimate of $c_{200}$ will be necessary for any stacking procedure if an unambiguous test of CDM is to be performed. Adopting a mass-concentration relation should be acceptable for rescaling real groups though. We have used exact concentrations for individual halos in this work as our aim was to test an idealized scenario to see if dynamical systematics may be significant. The primary effect of using a mass-concentration relation instead should be an increased uncertainty in the rescaled parameters, although further tests with more realistic data should of course be performed before any application to observations. We qualitatively discuss several avenues for determining the mass-concentration relation below.

The first would be to use the internal stellar dynamics of the centrals themselves. This route has a major physical hurdle and a major observational hurdle in the context of our work. On the physical side, the stellar dynamics are susceptible to baryonic feedback physics \citep[e.g.][where feedback can induce cored features]{Lazar_2020} and this introduces a systematic uncertainty that is difficult to quantify. On the observational side, this method would also require spatially resolved observations of the galaxy, which will not be available for the majority of centrals. This method is therefore not viable for a DESI-like survey.

Another option would be to use weak lensing observations of the groups to constrain the mass-concentration relation. Again there are theoretical and pragmatic issues with this route. If we aim to test for potential differences between a dynamical analysis and galaxy-galaxy weak lensing studies, then calibrating the dynamics using weak lensing data would undermine the point of the exercise. 

More indirect methods could also be considered. For example, \citet{Yang_2025} use weak lensing observations to measure the correlation between the central:satellite luminosity ratio and halo concentrations. This luminosity ratio should be simple to extract for any choice of redshift survey, but might induce a large uncertainty when converting from luminosity ratio to concentration. Indirect methods such as this may also be reliant on weak lensing measurements to calibrate them.

A more realistic method to `directly' measure the group concentrations could be through the satellite dynamics themselves. Since only the inner regions of the halo are required to estimate concentration, satellites in individual groups could be selected to be within the projected half-light (or half-mass) radius. This is the radius that encloses half of the total luminosity (mass) of the group and is typically about a quarter of the virial radius. Bias effects discussed in \autoref{subsec:correction terms} will be negligible at such small radii \citep[e.g.][]{Shi_2024}. The idea would then be to build a mass-concentration relation by measuring satellite velocity dispersions of stacked groups in real space and in narrow-mass bins. If this relation is unbiased and its scatter is accurately constrained, then this should be sufficient for our rescaling procedure. This method would be simple to implement as it would utilise the same data as the full dynamical mass profile measurement.

Obtaining unbiased estimates of a mass-concentration relation should be feasible with current data and using this relation should be adequate for our rescaling methodology. The main drawback we expect is that the recovered mass profiles will be less strongly constrained.

\subsubsection{Uncertainties on rescaling parameters}

Throughout this work we have assumed that $M_{200}$, $R_{200}$ and $c_{200}$ are known perfectly, but this is an unrealistic expectation with observational data. The confidence intervals presented here are therefore smaller than we would expect for real data. Groupfinders will be able to identify $M_{200}$ with an uncertainty of $\sim 0.2$ dex \citep{Tinker_2020} and the mass-concentration relation is expected to have an intrinsic scatter of $\sim0.09 - 0.13$ dex \citep{Umetsu_2020}. The velocity uncertainties of modern redshift surveys ($\sim10$ km/s) and the intrinsic scatters about the true median profile should be subdominant to these effects. The general methodology we present should therefore produce consistent results (although with inflated confidence intervals) so long as estimates on these parameters are unbiased and the scatter around the true values are understood. Two potential caveats we foresee are with the convolution of the velocity dispersion with the error distribution-function (\autoref{subsubsec:Methods vel disp calc}) and with the assumption of no uncertainty on the projected separations $\tilde{R}$.

For the convolution, we have used the analytic approximation that is inbuilt to the \textsc{binulator} code \citep{Gravsphere_binulator}. However this formula is inaccurate when the velocity uncertainties and velocity dispersions are a similar order of magnitude. This is likely to occur with more realistic observational uncertainties and will require a more involved deconvolution process. This is not an insurmountable issue, so long as the uncertainty PDFs are understood.

For the uncertainties on projected separations, an uncertainty on $c_{200}$ will introduce an uncertainty on the rescaled projected separation. In a worst case scenario, this could lead to high probabilities of data points actually lying in adjacent bins, especially for those already near a bin edge. This issue is also not impossible to solve, but will require a more involved and computationally expensive binning process to be developed. One solution could be to sample positions by a random draw from their PDFs for each MCMC model and then rebinning the data accordingly. This should lead to accurate velocity dispersion measurements that take these position uncertainties into account and can be passed on to \textsc{gravsphere}. Another option would be switching to a code that does not require binning of the data, such as the recently released Gravsphere2 \citep{Gravsphere2_2025}.

\subsubsection{Quality cuts}

Another area of concern could be the viability of applying the quality cuts explored here to real data. For example, the infall times of satellites are not directly measureable. However, since the quenched sample provided the most accurate dynamical mass profile of our \textit{Sat-} samples, we will assume for now that selecting for quenched satellites will be sufficient and that this quality cut will not prove problematic to implement.

\subsubsection{Modelling improvements}
A possible area of improvement would be to use a more appropriate model for the mass profiles. Whilst the cNFWt model (\autoref{subsubsec:Mass model}) is more flexible than NFW, it is still limited to a constant logarithmic density slope at large radii and only accommodates one transition radius. This might not accurately reflect stacked data. In addition, the numerical values of key parameters such as $M_{200}$ and $c_{200}$ have little physical significance in the rescaled space. It may be beneficial in the future to generalise this model and consider additional transition radii for increased flexibility.

\subsection{Future prospects}

\subsubsection{Application to current redshift surveys}

The methodology we present here could be applied to spectroscopic surveys that are sufficiently wide or deep to provide an adequate number of satellites that can be used to probe groups.

The DESI Bright Galaxy Survey \citep[DESI BGS;][]{Hahn_2023} should be viable for the wide case. It aims to measure 10s of millions of galaxy spectra at $z<0.6$ over five years, with the group catalogue for the first year of these data now publicly available \citep{Williams_2026}. Redshift uncertainties are small \citep[$\lesssim10$ km/s;][]{DESI_validation_2023} and it observes galaxies to stellar masses as low as $10^{7}M_{\odot}$. With such a survey it is reasonable to expect at least $100,000$s of satellites available to measure stacked dynamical mass profiles in galaxy groups down to at least halo masses of $M_{200}\sim10^{13}M_{\odot}$. This would constitute more than an order of magnitude lower in halo mass than previous dynamical studies on individual clusters \citep{Li_2023} and such an analysis is theoretically possible with the currently available data.

Another promising survey is the Wide Area Vista Extragalactic Survey \citep[WAVES;][]{WAVES_2019}, which is designed to take deeper observations specifically of galaxy groups and may allow stacked dynamical mass profiles to be measured down to masses of $M_{200}\sim 10^{11}M_{\odot}$, albeit with fewer tracers than will be available with DESI BGS. Such low masses would be comparable to the upper mass limit of the galaxy-galaxy weak lensing studies \citep{Brouwer_2021,Mistele_2024_RAR,Mistele_2024} and would fully bridge the gap between the galaxy scale and cluster scale. 

\subsubsection{Other applications}

Measuring velocity anisotropies is non-trivial in distant galaxy groups as only line-of-sight information is available. Whilst this is a nuisance parameter when trying to recover the mass profile, $\beta$ itself encodes information about the accretion of matter onto the halo \citep{Borzyszkowski_2017}. The cosmic web environment of a halo is also expected to modulate its accretion rate \citep{Tojeiro_2025} and so comparing the velocity anisotropies of samples within these different environments can allow for a test of the hierarchical structure formation in $\Lambda$CDM. However, \autoref{fig:beta normal} shows significant biases in the recovered velocity anisotropy profiles at large radii. Further work to understand and hopefully correct these biases would need to be performed before any application to data.

\section{Conclusions}\label{sec:conclusion}

In this work, we have outlined a new rescaling and stacking methodology to test the NFW mass profile of cold dark matter \citep{NFW1996,NFW_1997}. We have tested the viability of this method using idealised, line-of-sight data at $z=0$ in the TNG300 simulation \citep{TNG_2019}. We analysed galaxy groups in the mass range $10^{11.3}M_{\odot}/h \le M_{200} \le 10^{14}M_{\odot}/h$ and used subhalos as tracers to measure dynamical mass profiles with the Jeans equation. We performed this analysis using the code \textsc{gravsphere} \citep{Gravsphere_1}, within which we made minor modifications.

Dynamical mass profiles were measured for 7 different samples after interlopers had been removed \citep[with a method derived from][]{Hartog_1996}. The first 3 utilised the dark matter substructure whilst the remaining 4 were limited to `visible' satellite galaxies. We found the recovered mass profiles (\autoref{fig:Mass profile Gravsphere}) differed substantially when using the dark matter substructure versus the satellite galaxies, even when comparable quality cuts had been performed. As a result, the best performing quality cuts also differed between the two.

For the dark matter substructure, selecting subhalos which infell $\gtrsim 4.3$ Gyr ago onto their group provided the best recovery of the true mass profile, with an unbiased dynamical mass profile up to $\sim2.5$ times the median virial radius. For the satellite galaxies, selecting based on infall time instead led to a systematic underestimate of the true mass profile. Instead, the best recovery was found when selecting for quenched satellites, where the true profile was recovered within $1 \sigma$ up to $\sim 1.5$ times the median virial radius. However, the uncertainties appeared too large to reject an infinitely flat rotation curve scenario \citep[as observed with weak lensing around isolated lens galaxies;][]{Brouwer_2021,Mistele_2024,Mistele_2024_RAR} beyond $\sim 2 \sigma$. The constraining power may be improved with the $\sim$ order of magnitude increase in the number of tracers expected with modern surveys such as DESI BGS \citep{Hahn_2023}.

For the remaining samples, where no further quality cuts were performed or dynamically relaxed groups had been selected \citep[using the criteria of][]{Neto_2007}, the dynamical mass profiles were biased high.

We also found that recovered velocity anisotropies were biased towards negative values for all 7 of our subhalo samples beyond the median host virial radius (\autoref{fig:beta normal}). This could be caused by an incomplete removal of interlopers (which would bias the fourth line-of-sight velocity moments high) or due to dynamical systematics. We provide a mathematical argument in \autoref{subsubsec:tangential motion} that bulk tangential motion (e.g. from halo spin) will not bias results for randomly oriented stacked systems as any rotational support is actually absorbed into the velocity anisotropy of the stacked data. Other effects such as bulk radial motion and the two-halo term may still be present. We plan to quantify these dynamical systematics in future work.

To summarise, our results show that using the dynamical modelling of stacked galaxy groups as a test of cold dark matter predictions is a promising line of research. It is reassuring that selecting for quenched satellite galaxies provides a good recovery of the true mass profiles as these are easy to identify within observational data. Still, biases in the velocity anisotropies strongly motivate further exploration of these data to increase the range of applications of this method. We also highlight that the use of dark matter substructure within hydrodynamical simulations (or dark matter only simulations) may provide misleading conclusions as to what quality cuts are necessary and effective. Future work will aim to quantify dynamical systematic effects before applications to spectroscopic surveys such as DESI BGS or WAVES.

\begin{acknowledgments}
AR is supported by the Science and Technology Facilities Council (STFC) [studentship ST/X508780/1 and grant number ST/V000861/1]. RT is supported by STFC [grant number UKRI4102]. The authors would like to thank the TNG team for their continued work to keep the IllustisTNG simulations publicly accessible. The authors would also like to thank Dylan Nelson for providing access and aid through the TNG forums and with the TNG JupyterLab workspace.
\end{acknowledgments}



\bibliographystyle{aasjournalv7}
\bibliography{Stacked_groups} 






\end{document}